\documentclass[Journal,letterpaper,10pt,InsideFigs, NoLineNumbers]{ascelike-2019} 

\usepackage[utf8]{inputenc}
\usepackage[T1]{fontenc}
\usepackage{lmodern}
\usepackage{graphicx}
\usepackage[figurename=Fig.,labelfont=bf,labelsep=period]{caption}
\usepackage{subcaption}
\usepackage{amsmath}
\usepackage{txfonts}
\usepackage[hyphens]{url} 
\usepackage[colorlinks=false,citecolor=black,linkcolor=black]{hyperref}
\newcommand{\doi}[1]{\url{#1}}
\usepackage[authoryear]{natbib} 
\setcitestyle{authoryear,aysep={}} 
\usepackage{pgfplots}
\usepackage{float}

\usepackage{multirow}
\usepackage{longtable}
\usepackage{booktabs}
\usepackage{textcomp}

\usepackage{mathptmx}

\DeclareMathAlphabet{\mathcal}{OMS}{cmsy}{m}{n} 

\usepackage{amstext}
\usepackage{amssymb}
\usepackage{esint}

\usepackage{siunitx}
\ifdefined\unit\else
	\ifdefined\NewCommandCopy
		\NewCommandCopy\unit\si
	\else
		\NewDocumentCommand\unit{O{}m}{\si[#1]{#2}}
	\fi
\fi

\usepackage[colorinlistoftodos,prependcaption,textsize=tiny,color=red!40]{todonotes}

\providecommand{\tabularnewline}{\\}

\newcommand{\quotes}[1]{``#1''}

\usepackage{stackrel} 

\usepackage{longtable} 
\NameTag{Li, \today}
\begin{document}

\title{Time-resolving PIV measurements and modal analysis of turbulent flow in a bench-scale hydrodynamic separator}

\author[1]{Haochen Li, Ph.D., A.M.ASCE}
\affil[1]{Assistant Professor, Water Infrastructure Laboratory, Department of Civil and Environmental Engineering, University of Tennessee, Knoxville, Tennessee 37996, USA Email: hli111@utk.edu}
\maketitle

\begin{abstract}
Effective stormwater treatment infrastructures are crucial for mitigating the adverse effects of runoff on urban water quality. However, designing cost-effective treatment systems can be challenging due to complex turbulent flow dynamics. This study presents an in-depth analysis of turbulent flow in hydrodynamic separators (HS) using time-resolving, high-resolution particle image velocimetry (PIV) and modal decomposition techniques. The examination of interrogation window sizes on PIV measurements highlights a trade-off between spatial resolution and measurement uncertainty. Additionally, the impact of sampling frequencies and durations on the convergence of turbulence statistics, such as mean flow and Reynolds stress, is quantified. Results indicate that higher-order statistics require significantly larger sampling sizes (5x) to achieve the same level of statistical convergence as mean flow statistics. Leveraging proper orthogonal decomposition (POD) and spectral proper orthogonal decomposition (SPOD), dominant turbulence structures and coherent flow features are identified, providing a foundation for the development of reduced-order models (ROM). These ROMs demonstrate improved computational efficiency and have potential for real-time control applications and integration into larger drainage-scale simulations. Furthermore, the outcome of this study establishes a high-quality, open-source turbulent flow database for HS systems, offering the civil and environmental engineering community valuable resources to guide the development and application of computational fluid dynamics (CFD) tools. This study represents a critical step toward the ultimate goal of facilitating scientifically guided HS design and optimization.

\end{abstract}

\section{Introduction}
Stormwater infrastructures are a critical component of urban water management systems. It encompasses a range of facilities and practices designed to manage runoff generated by precipitation events, such as rain or snowmelt, in urban areas. As cities expand and impervious surfaces like roads, rooftops, and parking lots increase, the natural infiltration of water into the ground is reduced. This leads to higher volumes of surface runoff, which can cause flooding, erosion, and water pollution \citep{Dean2005, NationalResearchCouncil2008}. Under climate change, these issues are expected to intensify due to more frequent and severe precipitation events \citep{IPCC2022, Hathaway2023}.

Considerable efforts have been devoted to developing effective stormwater infrastructure solutions to control the flow and improve the quality of stormwater, mitigating the adverse effects on urban environments and water bodies \citep{Weiss2005}. These solutions include nature-based green infrastructure approaches such as rain gardens, green roofs, and permeable pavements \citep{Collins2008}, as well as engineered solutions such as detention basins \citep{Li2021basin212}, hydrodynamic separators \citep{Wilson2009, Li2020JEENJCAT}, and adsorptive filters \citep{Erickson2012, Li2022Adsorption}. Among these, hydrodynamic separators (HS) are commonly applied in densely populated areas where space is limited, and efficient pollutant removal from stormwater is required. HS utilizes sedimentation principles to separate various pollutants from stormwater, such as coarse sediments, floatables, and oil and grease \citep{Dickenson2009}. These systems are designed to be compact and efficient, making them suitable for urban environments where space constraints are a significant consideration. In the U.S., the development of HS systems is primarily driven by commercial interests, and their designs are often proprietary. Typically, these systems are physically tested in certified laboratories following standardized protocols, and their performance is reported and verified through third-party organizations such as the New Jersey Corporation for Advanced Technology (NJCAT) \citep{NJCAT2021, Committee2014}.

Designing an efficient HS system that enhances particulate matter (PM) separation and pollutant removal can be a challenging task. The system treatment performance results from complex interactions between turbulent flow and coupled pollutant transport \citep{Li2021Mechanica}. The chaotic nature of turbulence and the nonlinear response of system performance can challenge design engineers' intuitions, making it difficult to directly deduce an optimal HS geometry. Indeed, a recent study by \citet{Li2022SOR} based on 162 publicly available physical testing results for 22 HS systems revealed that commercial HS systems, despite their complex and proprietary designs, do not demonstrate more effective designs than plain cylinder tanks for PM separation \citep{Howard2012}. These findings highlight the critical need for improved design strategies for HS systems.

Conceptually, HS system design is an optimization and inverse problem, where the goal is to identify the HS system geometries that maximize certain user-defined objective functions such as PM separation \citep{Li2022basinOpt}. This inverse identification process can be iteratively performed through either physical modeling or numerical modeling. Compared to physical modeling, which requires design engineers to repeat the physical testing for each system design alteration, a numerical modeling approach based on computational fluid dynamics (CFD) can simulate the HS system hydrodynamics and predict pollutant transport and fate, obviating the need for time-consuming physical testing and reducing the overall capital investment and risk in HS product development. As computational hardware advances and CFD models progress, there are increasing applications and significant potentials for using CFD to guide next-generation HS system design and optimization \citep{Li2020JEENJCAT, Li2020JEEresuspension}.

The success of CFD in guiding HS system design relies on its predictive capability \citep{Liu2020a}. CFD models, particularly those utilizing turbulence closures such as Reynolds-Averaged Navier-Stokes (RANS) equations, often need to be validated against experimental data to ensure the validity of the underlying turbulence models and the robustness of predictions of system hydrodynamics. Reliable experimental data is crucial for benchmarking and validating these CFD models \citep{Nopens2020, Catano-Lopera2023}. In the existing literature, several experimental measurements for HS systems have been undertaken. For instance, \citet{Pathapati2009a} performed PM separation tests of polydispersed particle size distributions for a full-scale HS with two cylindrical chambers. \citet{Howard2012,Howard2010thesis} conducted PM separation and acoustic Doppler velocimetry (ADV) flow measurements for two commercial HS systems and a cylindrical tank. These studies primarily focused on the treatment performance of HS systems at full scale. While full-scale experiments provide analogs to in-situ conditions, they yield limited spatial-temporal resolutions and insights into flow physics. The HS systems examined in these studies were also proprietary (except for the cylindrical tank), with confidential details on their geometries. In recent years, researchers have applied particle image velocimetry (PIV) for flow measurements in bench-scale HS models or systems with similar geometries. \citet{Beg2018} conducted stereo PIV measurements for a bench-scale manhole, while \citet{Wang2023b} performed 2D planar PIV on bench-scale cylindrical tanks and baffled cylindrical tanks. Compared to point measurement techniques like ADV, PIV provides significantly enhanced spatial resolution, allowing flow velocity to be measured across a 2D plane or even within a 3D volume (i.e., tomographic PIV). A close integration of these pioneering research results indicates that these databases can be further improved by (1) providing detailed experimental procedures and documentation for reproducibility, (2) increasing spatial-temporal resolution, (3) examining statistical convergence in mean flow and turbulence quantities (e.g., insufficient convergence can often appear as non-smoothness in flow statistics contours), and (4) enhancing visualization of flow physics.

This study presents high-resolution, time-resolved PIV measurements of turbulent flow in a bench-scale HS. The motivation for this research is twofold: first, to elucidate the flow hydrodynamics and turbulence statistics within an HS system, and second, to develop a high-quality open-source database that enables engineers to benchmark and validate their CFD models. This validation process is crucial for assessing the predictive capability of CFD models and examining the sensitivity of numerical configurations (e.g., mesh, scheme, and linear solver configuration). More importantly, reproducible experiment data and benchmark cases can provide engineers with confidence in CFD technology, promoting its adoption in civil and environmental applications \citep{Catano-Lopera2023}. The specific objectives of this study are: (1) to examine the convergence of turbulence statistics with respect to the sampling number and sampling frequency; (2) to elucidate the mean flow and Reynolds stress distribution within the HS system; and (3) to investigate the spectral and modal characteristics of the turbulent flow.

\section{Methodology}

\subsection{Bench-scale hydrodynamic separator geometry}
\begin{figure}[H]
\centering
\includegraphics[width=0.8\textwidth]{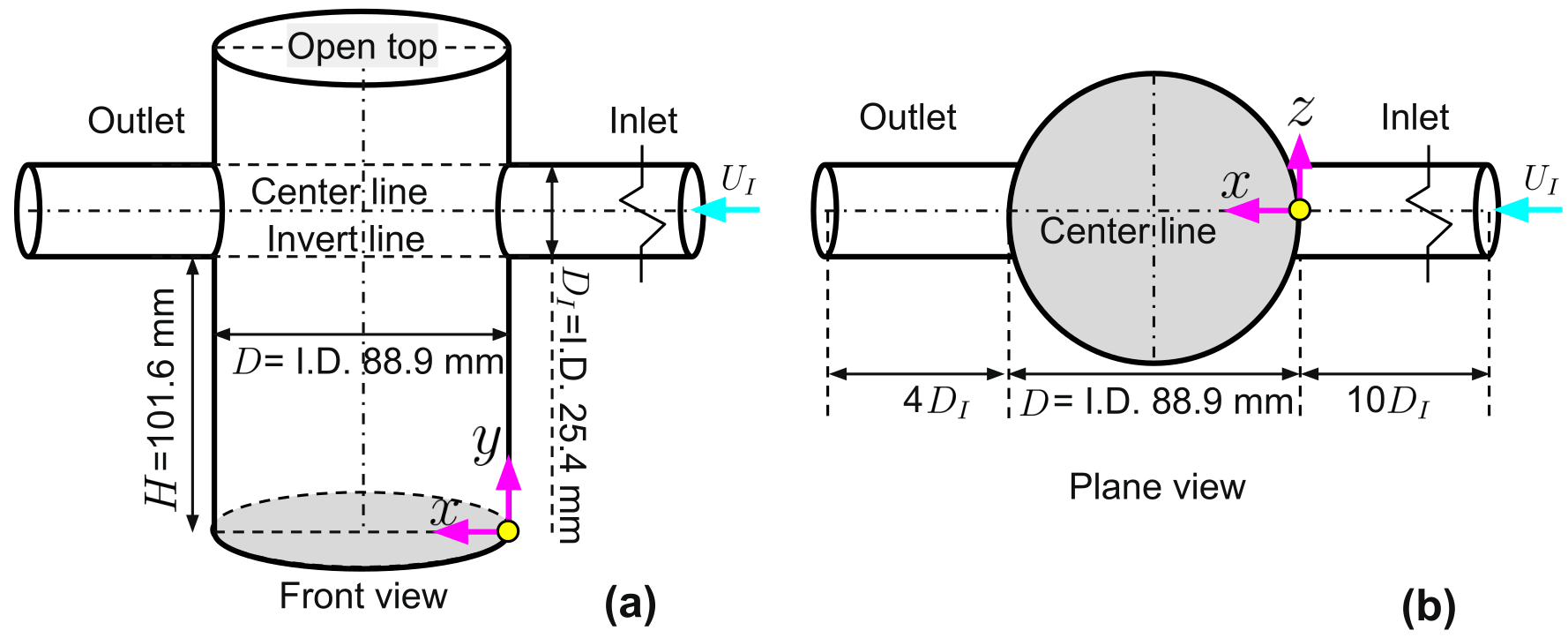}
\caption{Geometric illustration of bench-scale hydrodynamic separator tested in Water Infrastructure Laboratory, University of Tennessee. (a) front view and (b) plane view. $D$ and $D_I$ are the inner diameter of the cylindrical tank and inlet pipe, $U_I$ is inlet bulk velocity. The experiment configuration is illustrated in Fig. \ref{fig:PIV_experiment}. The tested flow conditions are summarized in Table \ref{tab:flow_conditions}.}
\label{fig:tank_geo}
\end{figure}

The physical modeling of a bench-scale HS made of transparent Plexiglas was conducted in the Water Infrastructure Laboratory at the University of Tennessee. Fig. \ref{fig:tank_geo} illustrates the geometry of the bench-scale HS. The geometric layout of the HS closely resembles a plain cylindrical tank. Specifically, the cylindrical tank comprises a large cylindrical volume with an inner diameter $D$ of \SI{88.9}{mm} (3.5 inches). The inlet and outlet pipes are configured along the system centerline with the same inner diameter $D_I$ of \SI{25.4}{mm} (1 inch). The system is symmetric with respect to the center plane ($z=0$), where the z-axis is perpendicular to the paper. This HS geometry was determined by considering and balancing several factors such as geometric similarity, PIV camera resolution, seeding particle Stokes number, and laser sheet divergence. The geometry of the HS system is designed with a depth-diameter ratio of 1.14, within the typical range of 1.0-1.23 for existing commercially available cylindrical HS systems \citep{Li2020JEENJCAT}. The bench-scale HS also has a similar geometry to a full-scale system tested in the St. Anthony Falls Laboratory (SAFL) at the University of Minnesota \citep{Howard2010thesis}.

This plain cylindrical geometry was selected for several reasons: (1) It represents the simplest elementary geometry for an HS, ensuring reproducibility. (2) These cylindrical tanks are commonly constructed at junctions or vertical chambers (\quotes{manholes}) of urban stormwater or wastewater conveyance networks for flow conveyance and PM separation. (3) As previously mentioned, commercial systems with more complex geometries do not demonstrate improved PM separation performance compared to plain cylindrical tank geometry \citep{Li2022SOR}. It is also worth noting that configuring a vertical baffle within the HS does not necessarily improve the system's PM separation. Conversely, both experiments and large-eddy simulations (LES) \citep{Li2021JFE} show that the unbaffled system can yield higher PM separation. (4) Despite their simple geometries, these cylindrical tanks can be subject to complex turbulent features and flow dynamics, including shear layers and jet impingement, as indicated by high-fidelity LES simulations \citep{Li2022SOR, Li2021Mechanica}. Therefore, understanding the dynamics of plain cylindrical tank geometry is particularly critical and can serve as a baseline comparison to any future HS design.

\subsection{Particle image velocimetry experiment configuration}
\begin{figure}[H]
    \centering
    \includegraphics[width=1.0\textwidth]{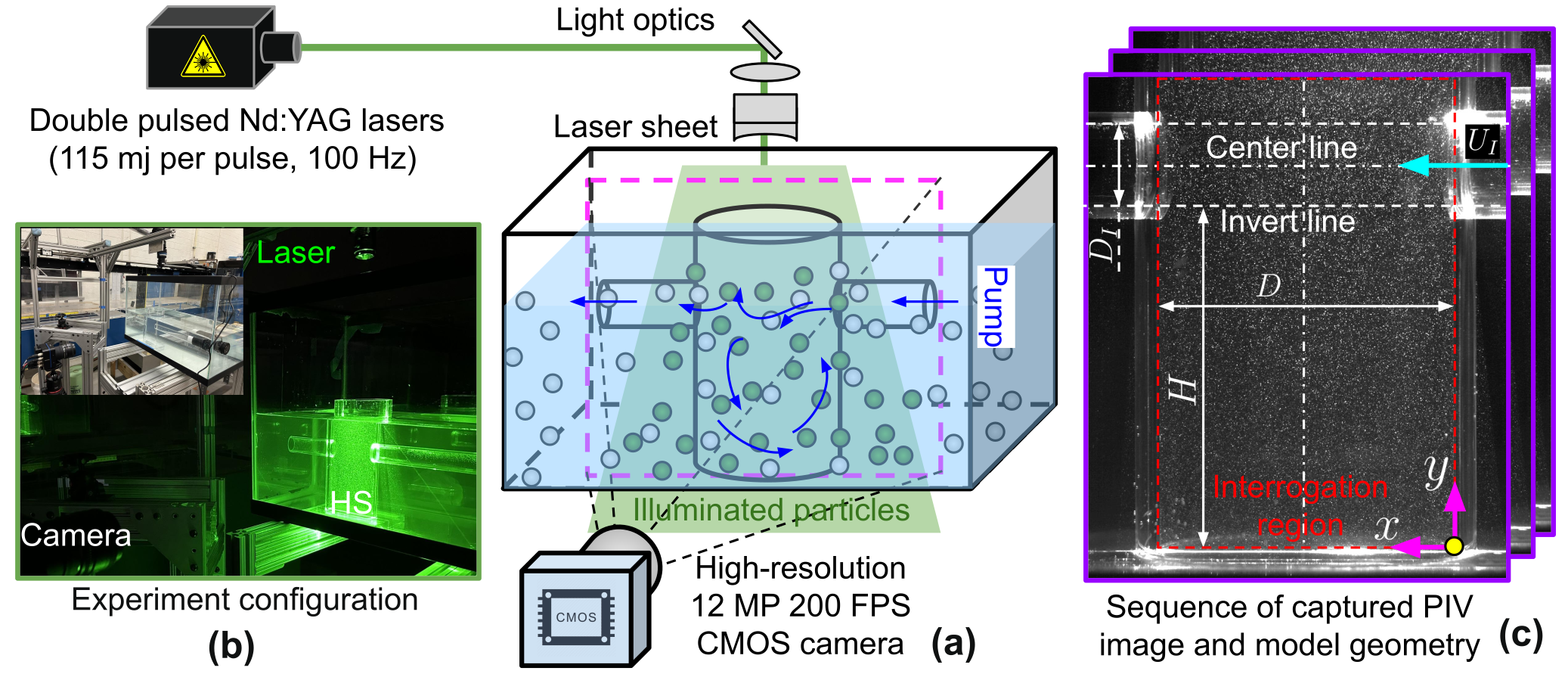}
    \caption{PIV experimental configurations bench-scale cylindrical tank in Water Infrastructure Laboratory at University of Tennessee. (a) schematic of the experimental setup; (b) visualization of the experiment; (c) example of captured PIV images. The geometric dimensions are shown in Fig. \ref{fig:tank_geo}. The tested flow conditions are summarized in Table \ref{tab:flow_conditions}.}
    \label{fig:PIV_experiment}
\end{figure}

Fig. \ref{fig:PIV_experiment} illustrates the 2D planar PIV experiment configuration. The bench-scale HS is installed in a transparent rectangular glass container, with the inlet connected to a submersible pump (Jebao SOW-3, maximum flow rate of 21 gallons per minute). The use of a rectangular glass container minimizes optical distortion caused by the curved walls of the HS model. With this configuration, optical distortion is controlled within a 10\% range, as shown in Fig. S1. The inlet pipe is extended to a total length of 10$D_I$ to facilitate the development of turbulence \citep{Cengel2013}. A \SI{25.4}{mm} (1 inch) long aluminum honeycomb grid with a cell size of \SI{3.175}{mm} (1/8 inch) is configured in a customized 3D-printed inlet fitting to mitigate the rotational effects of the pump propeller and expedite turbulence reequilibration in the inlet pipe. The transparent rectangular glass container, along with the bench-scale HS model, is filled with distilled water to a water elevation of 1.5$D_I$ above the inlet centerline at \SI{20}{\degreeCelsius} (fluid density $\rho_f$ = \SI{998}{kg/m^3}, kinematic viscosity $\nu$ = \SI{1e-6}{m^2/s}).

A dual-head 100 Hz Nd:YAG laser (Litron, LPY704-100, \SI{532}{\nm}, 115 mJ/pulse) is used as the laser source, and an optical arm and a series of light optics (spherical lens and cylindrical lens) are used to deliver a 2D laser sheet from the top of the HS model. A laser sheet with a thickness of \SI{1}{mm} is generated and aligns with the center plane of the HS model ($z$ = 0). A 12-MP CMOS camera (Vieworks, VC-12MX-18, 4096$\times$3072 pixels, 180 fps, pixel pitch of \SI{5.5}{\um}) with a 50 mm lens (Nikon AF Nikkor, f/1.8D) is positioned orthogonal to the laser sheet. The portion of the field of view is illustrated in Fig. \ref{fig:PIV_experiment}c. The images are acquired through an imaging software platform, TSI InsightV3V™ 4G, which controls a TSI synchronizer (Model 610036) that simultaneously triggers and stops the camera and the pulsed laser. Double images with an interframe time of $\Delta t=$ \SI{1}{\ms} are acquired at 50 Hz. The interframe time is selected such that the maximum particle displacement is approximately 25\% of the size of the interrogation window (32$\times$32 pixels) at the inlet region. 

Distilled water is mixed with \SI{10}{\um} neutrally-buoyant glass hollow sphere seeding particles and added to the rectangular container and HS model. The seeding particles have a particle relaxation time scale $\tau_p$ of \SI{5.6}{\us}, as defined in Eq. \ref{eq:particle_time}. The Kolmogorov length scale $\eta$, and time scale $\tau_k$ of flow turbulence are estimated to be \SI{30}{\um} and \SI{0.87}{\ms}, respectively, as defined in Eqs. \ref{eq:Kolmogorov_length} and \ref{eq:Kolmogorov_time}, calculated at the inlet jet for the highest flow rate case \citep{Balachandar2010}. The particle Stokes number $St$ is therefore approximately \num{6.3e-3}, as detailed in Eq. \ref{eq:particle_stokes}. This low particle Stokes number indicates that the seeding particles are sufficiently small to closely follow the turbulent flow, ensuring accurate PIV measurements \citep{Balachandar2010, ExperimentalFluidMechanics2007}. The seeding density of the glass hollow spheres is adjusted and optimized in preliminary experiments so that each local interrogation window (32$\times$32 pixels) contains 5-15 particles, ensuring robust PIV measurement \citep{Scharnowski2020a}. Camera calibration is performed with a \SI{2}{\mm} $\times$ \SI{2}{\mm} checkerboard aligned with the laser sheet at the measurement center plane, as shown in Fig. S1. The 2D planar PIV measurements are conducted for four different flow rates. These rates are set to turbulent flow regime and to yield a similar surface loading rate (SLR) in the bench-scale HS to the common SLR range of 1020-1824 \unit{L/min/m^2} found in existing commercially available HS systems \citep{Li2020JEENJCAT, Li2022SOR}. Table \ref{tab:flow_conditions} summarizes the information for the different cases.

\begin{equation}
    \tau_p = \frac{\rho_p}{\rho_f}\frac{ d_p^2}{18\nu}, \label{eq:particle_time}
\end{equation}

\begin{equation}
    \eta = \frac{D_{I}}{{Re}_I^{3/4}}, \label{eq:Kolmogorov_length}
\end{equation}

\begin{equation}
    \tau_k = \frac{\eta^2}{\nu}, \label{eq:Kolmogorov_time}
\end{equation}

\begin{equation}
    St = \frac{\tau_s}{\tau_k}. \label{eq:particle_stokes}
\end{equation}
In these equations, $\tau_p$ is particle relaxation time scale, $\eta$ and $\tau_k$ are the Kolmogorov length scale and time scale, $St$ is the particle Stokes number, and ${Re}_I$ is inlet Reynolds number.

\begin{table}[H]
\caption{Summary of PIV experiment configurations. ($U_{I}$ is inlet bulk velocity; $Q$ is inlet flowrate; SLR$=Q/A$ is surface loading rate; ${Re}_I = U_ID_I/\nu$ is inlet Reynolds number; $Re_{b} = \mathrm{SLR}\times D/\nu $ is bulk Reynolds number; ${Fr}_b = U/\sqrt{gH}$ is bulk Froude number; $t_{c} = D_I/U_I$ is convective time unit; $T_d$ is sampling duration; $A$ is HS system bottom area; $g$ is gravitational acceleration.) \label{tab:flow_conditions}}
\centering{}%
\setlength\tabcolsep{4pt} 
\begin{tabular}{ccccccccc}
\toprule 
Case & $U_{I}$ & $Q$ & SLR & ${Re}_{I}$ & ${Re}_{b}$ & ${Fr}_b$ & $t_c$ & $T_d$ \tabularnewline
 \# & mm/s & L/min & \unit{L/min/m^2} & / & / & / & s & $t_c$ \tabularnewline
\midrule
1 & 249.7 & 7.6 & 1193.3 & 6342 & 1789 & 0.25 & 0.102 & 1769.5 \tabularnewline
2 & 282.9 & 8.6 & 1352.0 & 7185 & 2027 & 0.28 & 0.090 & 2004.8 \tabularnewline
3 & 304.8 & 9.3 & 1456.6 & 7741 & 2184 & 0.30 & 0.083 & 2160.0 \tabularnewline
4 & 321.7 & 9.8 & 1537.4 & 8170 & 2306 & 0.32 & 0.079 & 2279.8 \tabularnewline
\bottomrule
\end{tabular}
\end{table}

\subsection{Flow data computation and turbulence statistics quantification}
For each PIV measurement, 9,000 paired images are acquired after the flow has reached a stationary state. A multi-pass PIV algorithm, the Recursive Nyquist Grid, is employed with an initial interrogation window size of 64$\times$64 pixels, which is refined to 32$\times$32 pixels to adaptively respond to local flow characteristics. Alternative interrogation window sizes of 64$\times$64 and 16$\times$16 pixels are also examined, as shown in the Results section. This processing technique enhances spatial resolution in areas with complex flow patterns while maintaining an adequate dynamic range. Between each pass, local vector validation based on the universal median test is applied to identify and correct spurious vectors through statistical validation and consistency checks. Vector conditioning is employed to interpolate missing vectors by recursively filling with the local mean value \citep{TSI2020}. In both the local vector validation and conditioning processes, a 5$\times$5 pixel spatial neighborhood is utilized. The wall time for processing each case is approximately 14 hours, utilizing dual Intel Xeon Gold 6130 CPUs (2.10 GHz, 32 cores).

Based on the PIV algorithm, the instantaneous $x$-axis and $y$-axis velocity fields, $u(x,y,t)$ and $v(x,y,t)$, are computed with a spatial resolution of 108$\times$178 (using 32$\times$32 pixels window size) and a temporal resolution of 50 Hz. Stationary flow statistics are computed by averaging over time. These averages are denoted by $\left\langle \cdot \right\rangle$ and are also referred to as the mean quantities. A prime symbol $\left(\cdot\right)^\prime$ represents a perturbation from the mean. Under this notation, $\left\langle u \right\rangle$ is the averaged streamwise velocity, $\left\langle v \right\rangle$ is the averaged vertical velocity, and $\left\langle u^\prime v^\prime \right\rangle$ is the Reynolds (shear) stress, as defined in Eq. \ref{eq:average-example} and \ref{eq:average-RSS-example}.

\begin{equation}
\left\langle{u}\right\rangle\left(x, y\right)=\frac{1}{T}\int_{T} u\left(x,y,t\right)\,dt, \label{eq:average-example}
\end{equation}

\begin{equation}
{u}^{\prime} \left(x, y, t\right) = {u}\left(x, y, t\right)-\left\langle{u}\right\rangle\left(x, y\right), \label{eq:average-fluc-example}
\end{equation}

\begin{equation}
\left\langle {u}^{\prime}{v}^{\prime}\right\rangle \left(x, y\right) =\frac{1}{T}\int_{T} u^{\prime}\left(x,y,t\right)v^{\prime}\left(x,y,t\right)\,dt. \label{eq:average-RSS-example}
\end{equation}

Quantification of differences between two different statistically averaged solutions $\langle \psi_1 \rangle$ and $\langle \psi_2 \rangle$ is often needed. We define the relative mean difference (RMD) in Eq. \ref{eq:RMD}.

\begin{equation}
\textrm{RMD}=\frac{\mathop{\mathbb{E}}\left[\left|\langle \psi_1 \rangle -\langle \psi_2 \rangle \right|\right]}{\mathop{\mathbb{E}}\left[\left|\langle \psi_2 \rangle \right|\right]}.\label{eq:RMD}
\end{equation}
In this equation, the definition of the operator $\mathop{\mathbb{E}}$ denotes the average over the entire domain. $\left| \cdot \right|$ denotes the absolute value.

\subsection{Proper orthogonal decomposition}
Proper orthogonal decomposition (POD) is a data-driven decomposition technique that seeks to partition a dataset as a linear combination of elementary contributions called modes \citep{BruntonKutz2019}. Using POD to decompose a turbulence field facilitates the extraction of the most energetic spatial flow structures from a given flow dataset. This allows us to identify and quantify the dominant modes and coherent turbulent structures of the flow and their associated energy content. The theoretical background and detailed methodology for POD analysis can be found elsewhere \citep{BruntonKutz2019, Taira2017}. Herein, only brief steps are illustrated. The flow instantaneous velocities from PIV are first collected and organized as a series of $m$ snapshots (in this case, 9,000). Each snapshot is a column vector of length $n$ (number of spatial points, $108\times178 = 19,224$ in this case). These snapshots are arranged into a matrix $\mathbf{X} \in \mathbb{R}^{n \times m}$, as defined in Eq. \ref{eq:data_layout}.

\begin{equation}
   \mathbf{X} = \begin{pmatrix}
   | & | & & | \\
   \mathbf{x}_1 & \mathbf{x}_2 & \cdots & \mathbf{x}_m \\
   | & | & & |
   \end{pmatrix}, \label{eq:data_layout}
\end{equation}
where $\mathbf{x} \in \mathbb{R}^{n \times 1}$ is the flow instantaneous velocity reshaped as a column vector, $n$ is the total number of spatial measurements, and $m$ is the number of snapshots in time. Next, the mean of the snapshots is subtracted from each snapshot to obtain the fluctuation matrix $\mathbf{X}' \in \mathbb{R}^{n \times m}$, as defined in Eq. \ref{eq:fluctuation}.

\begin{equation}
   \mathbf{X}' = \mathbf{X} - \overline{\mathbf{x}} \mathbf{1}_m, \label{eq:fluctuation}
\end{equation}
where the mean field $\overline{\mathbf{x}} \in \mathbb{R}^{n \times 1}$ is the time-averaged vector, calculated as the average of each spatial point across all snapshots, as defined in Eq. \ref{eq:POD_time_average}. $\mathbf{1}_m \in \mathbb{R}^{1 \times m}$ is a column vector of ones with length $m$.

\begin{equation}
    \overline{\mathbf{x}} = \frac{1}{m} \sum_{j=1}^{m} \mathbf{x}_j, \label{eq:POD_time_average}
\end{equation}
Where $\mathbf{x}_j$ is the $j$-th snapshot. The covariance matrix $\mathbf{C} \in \mathbb{R}^{n \times n}$ is then constructed with the fluctuation matrix, as shown in Eq. \ref{eq:POD_covariance_matrix}.

\begin{equation}
    \mathbf{C} = \frac{1}{m} \mathbf{X}' \mathbf{X}'^\top, \label{eq:POD_covariance_matrix}
\end{equation}
Subsequently, eigenvalue decomposition is performed over the covariance matrix to obtain the matrix of eigenvectors $\boldsymbol{\Phi} \in \mathbb{R}^{n \times n}$ and a diagonal matrix of eigenvalues $\boldsymbol{\Lambda} \in \mathbb{R}^{n \times n}$, as defined in Eqs. \ref{eq:POD_eigen_decomp}-\ref{eq:POD_eigen_value_matrix}.

\begin{equation}
    \mathbf{C} \boldsymbol{\Phi} = \boldsymbol{\Phi} \boldsymbol{\Lambda}, \label{eq:POD_eigen_decomp}
\end{equation}

\begin{equation}
    \boldsymbol{\Phi} = \begin{pmatrix}
   | & | & & | \\
   \boldsymbol{\phi}_1 & \boldsymbol{\phi}_2 & \cdots & \boldsymbol{\phi}_n \\
   | & | & & |
   \end{pmatrix},
\end{equation}

\begin{equation}
\boldsymbol{\Lambda} = \mathrm{diag}(\lambda_1, \lambda_2, \lambda_3, \ldots, \lambda_n), \label{eq:POD_eigen_value_matrix}
\end{equation}
where each column of the eigenvector matrix $\boldsymbol{\Phi}$, denoted as $\boldsymbol{\phi}_i \in \mathbb{R}^{n \times 1}$, represents the $i$-th POD mode, and the eigenvalues $\lambda_i$ indicate the energy associated with the $i$-th mode.

The projection coefficient matrix $\mathbf{A} \in \mathbb{R}^{n \times m}$ can be obtained by projecting the original data onto the POD modes, as defined in Eq. \ref{eq:POD_A_projection}.

\begin{equation}
    \mathbf{A} = \boldsymbol{\Phi}^\top \mathbf{X}'. \label{eq:POD_A_projection}
\end{equation}
The interpretation of the projection coefficient matrix $\mathbf{A}$ provides insights into the contribution of POD modes across different snapshots. Examining the matrix by columns, i.e., $\mathbf{A} = [\mathbf{a}_1 \, \mathbf{a}_2 \, \cdots \mathbf{a}_m]$, each $\mathbf{a}_j$ (the $j$-th column of $\mathbf{A}$) is a vector of coefficients for all modes in the $j$-th snapshot. Conversely, examining the matrix by rows, each $i$-th row $\mathbf{r}_i = (a_{i1} \, a_{i2}\, \cdots \, a_{im})$ contains the coefficients for the $i$-th POD mode across all snapshots. The coefficients in a row provide the temporal dynamics, showing how the contribution of a particular POD mode varies over time (with snapshots organized by time). The POD method presented above is often referred to as direct POD. Alternatively, POD can also be carried out through the temporal covariance matrix for faster computation, which is termed snapshot POD. Another variation of POD can be conducted through singular value decomposition (SVD). The connections between these different approaches are discussed elsewhere \citep{Weiss2019, BruntonKutz2019}.

\subsection{Spectral proper orthogonal decomposition}
The principle of spectral proper orthogonal decomposition (SPOD) shares similarities with POD \citep{Lumey2007, Schmidt2020a}. However, instead of directly applying eigen decomposition in the time domain as in POD, SPOD operates the eigen decomposition in the frequency domain. Furthermore, because of the use of the discrete Fourier transform (DFT) to isolate each frequency, SPOD modes are not only coherent in space but also in time. In contrast, POD only ensures that the computed modes are coherent in space. SPOD is particularly advantageous in analyzing statistically stationary turbulent flows and represents an optimally averaged dynamic mode decomposition (DMD) obtained from an ensemble DMD \citep{Towne2018}.

The general procedure for SPOD is as follows. The fluctuation velocity field data $\mathbf{X}^\prime \in \mathbb{R}^{n \times m}$ is first segmented into $N_b$ overlapping time blocks. This segmentation process is in the spirit of the Welch method \citep{Welch1967} and aims to reduce the variance of the spectral estimate by averaging multiple blocks of the snapshots, as shown in the following section. Each block denoted as $\mathbf{X}^\prime_k \in \mathbb{R}^{n \times b}$, contains $b$ snapshots and represents a portion of the total snapshots, where $k$ is the block index. Then, a windowing function $\mathbf{W}$ (a Hamming window is used in this case) is applied to each block to reduce spectral leakage \citep{Schmidt2020a}, as defined in Eq. \ref{eq:SPOD_windowing}.

\begin{equation}
    \mathbf{X}_k^w = \mathbf{X}^\prime_k \circ \mathbf{W}, \label{eq:SPOD_windowing}
\end{equation}
where $\mathbf{W} \in \mathbb{R}^{n \times b} $ and $\circ $ denotes element-wise multiplication. Each windowed block is then transformed from the time domain to the frequency domain using the DFT. This results in a series of Fourier coefficients $\hat{\mathbf{X}}_k \in \mathbb{C}^{n \times b}$ for each block, as defined in Eq. \ref{eq:SPOD_Fourier_coefficients}.

\begin{equation}
    \hat{\mathbf{X}}_k = \mathcal{F}(\mathbf{X}_k^w), \label{eq:SPOD_Fourier_coefficients}
\end{equation}
where $\mathcal{F}$ denotes the DFT. For each frequency $f$, a cross-spectral density (CSD) matrix is computed. The CSD matrix $\mathbf{S}(f) \in \mathbb{C}^{n \times n}$ is obtained by averaging over all blocks, as defined in Eq. \ref{eq:SPOD_CSD}. The CSD matrix captures the spatial correlations of the velocity field at that particular frequency. 

\begin{equation}
    \mathbf{S}(f) = \frac{1}{N_b} \sum_{k=1}^{N_b} \hat{\mathbf{X}}_k(f) \hat{\mathbf{X}}_k^H(f), \label{eq:SPOD_CSD}
\end{equation}
where \(\hat{\mathbf{X}}_k(f)\) is the DFT coefficient at frequency $f$ for the $k$-th block, and \(H\) denotes the Hermitian (complex conjugate) transpose. The eigenvalue decomposition is then performed for the CSD matrix \(\mathbf{S}(f)\) at each frequency to obtain the eigenvalues $\boldsymbol{\Lambda}(f)$ and eigenvectors $\mathbf{V}(f)$, as shown in Eq. \ref{eq:SPOD_eigen_decomp}.

\begin{equation}
    \mathbf{S}(f) \mathbf{V}(f) = \mathbf{V}(f) \boldsymbol{\Lambda}(f). \label{eq:SPOD_eigen_decomp}
\end{equation}
In this equation, $\mathbf{V}(f) \in \mathbb{C}^{n \times n}$ is the eigenvectors, representing the SPOD modes. $\boldsymbol{\Lambda}(f) \in \mathbb{R}^{n \times n}$ is the eigenvalues indicating the energy associated with each mode at the given frequency. The SPOD modes are ranked based on their eigenvalues at each frequency, providing a frequency-dependent decomposition of the flow field, i.e., SPOD modes are orthogonal under a space-time inner product \citep{Schmidt2020a}. For real-valued signals such as flow velocity, only half of the spectrum is needed to represent the full signal, and Fourier coefficients are corrected for one-sided spectrum \citep{Lumey2007,Pope2000}.

\section{Results}
\subsection{Effects of PIV processing window size on flow statistics and uncertainty}
\begin{figure}[H]
    \centering
    \includegraphics[width=0.95\linewidth]{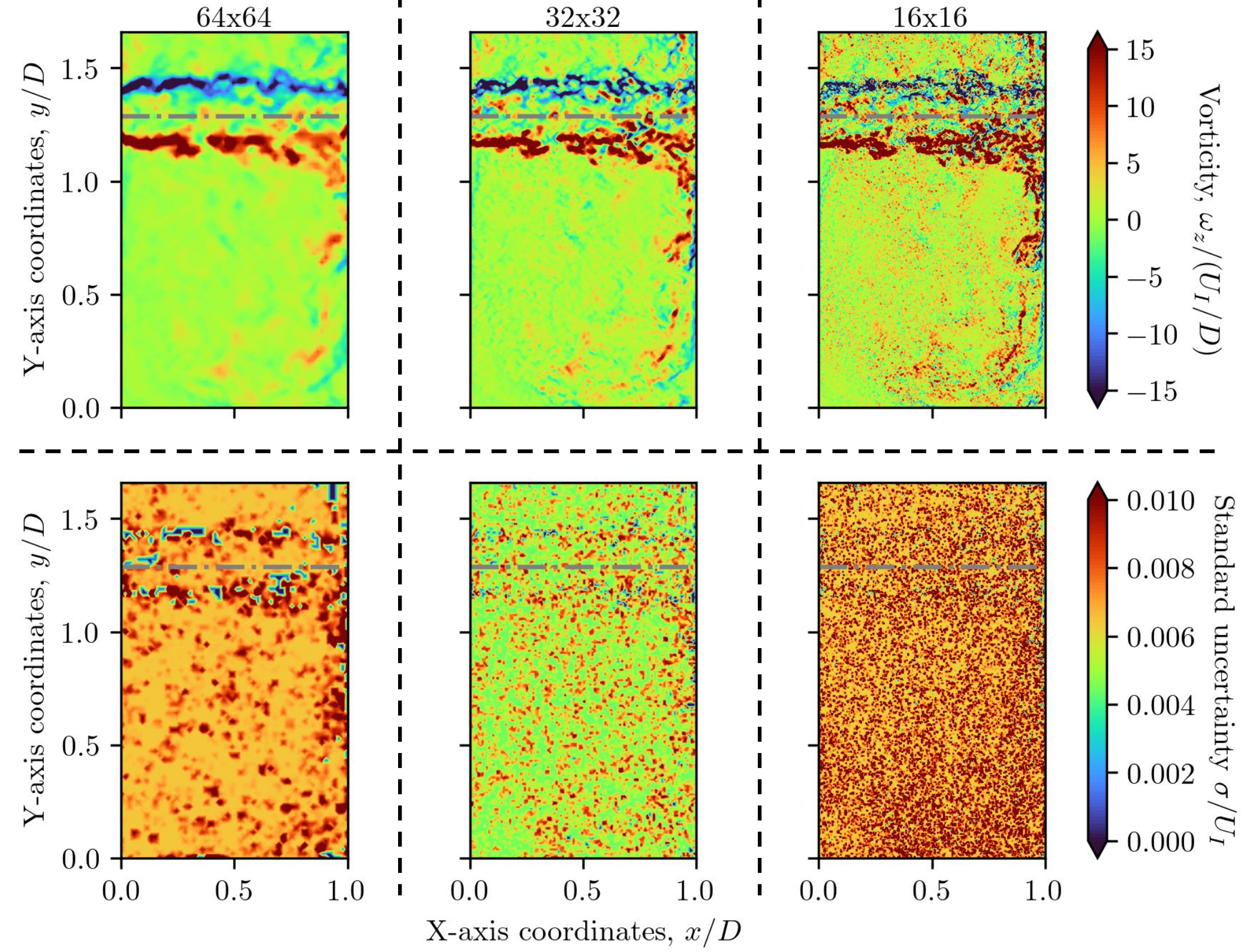}
    \caption{Z-axis vorticity and standard uncertainty with different window sizes. From left to right, each column corresponds to a window size of 64$\times$64, 32$\times$32, and 16$\times$16 pixel. $\omega_z$ is z-axis vorticity. $U_I$ is inlet bulk velocity, and $D$ is the inner diameter of cylinder. PIV measurements for case 4 are used. The standard uncertainty $\sigma$ is estimated based on the peak ratio method developed by \citet{Gauthier2013}.}
    \label{fig:PIV_window_size}
\end{figure}

Fig. \ref{fig:PIV_window_size} illustrates instances of computed z-axis vorticity contours $\omega_z$ using different PIV processing window sizes: 64$\times$64, 32$\times$32, and 16$\times$16 pixels. The 16$\times$16 pixels window, being the highest resolution, displays the most detailed structures in the vorticity field. Turbulent structures are clearly visible, particularly in the upper region. The intermediate resolution of the 32$\times$32 pixel window captures primary flow structures but with less detail compared to the 16$\times$16 pixel window. While the overall patterns of vorticity remain similar, some finer details are lost. The 64$\times$64 pixel window, the lowest resolution, shows a coarser representation of the vorticity field. Many of the finer turbulent structures are no longer visible, and the vorticity patterns appear more smoothed out.  The overall trend shows that as the window size increases, the spatial resolution decreases, leading to a smoother and less detailed depiction of the vorticity field. These effects of window size can also extend to the flow statistics to a degree. Generally, the mean flow statistics are less influenced by the window size, whereas larger window sizes tend to yield lower Reynolds stress, as shown in Figs. S2-5 (online Supplemental Material).

Notwithstanding, using a smaller processing window can increase measurement uncertainty. Fig. \ref{fig:PIV_window_size} presents the normalized standard uncertainty estimate $\sigma$ based on the empirical method developed by \citet{Gauthier2013}. This method uses the peak ratio between the largest detectable correlation peak and the second-highest peak. It provides a bulk estimation of uncertainty including many sources, such as seeding density, particle size, and velocity gradients, etc. These standard uncertainty results indicate that the smaller window size 16$\times$16 while enhancing resolution, increases measurement uncertainty by approximately 30\%. Larger process window sizes of  64$\times$64 also yield comparable uncertainty to the 16$\times$16 window. The 32$\times$32 window, showing the least standard uncertainty, is thus used for subsequent analyses.

\subsection{Convergence of mean flow statistics and Reynolds stress}
Robust computation of turbulence statistics such as mean flow and Reynolds stresses, relies on the statistical convergence of sampled data. Without this convergence, these statistics may be biased or inaccurate, a concern that was not diligently reported in some prior studies. Therefore, this study investigates the dependency of statistical convergence on the sampling frequency ($f_s$) and sampling duration ($T_d$) for first (mean flow) and second-order turbulence statistics (Reynolds stress). The convergence is quantified using the relative mean difference (RMD), defined in Eq. \ref{eq:RMD}, with respect to the reference statistics from the most conservative scenario, i.e., 50 Hz for \SI{180}{s}.

\begin{figure}[H]
    \centering
    \includegraphics[width=1.0\textwidth]{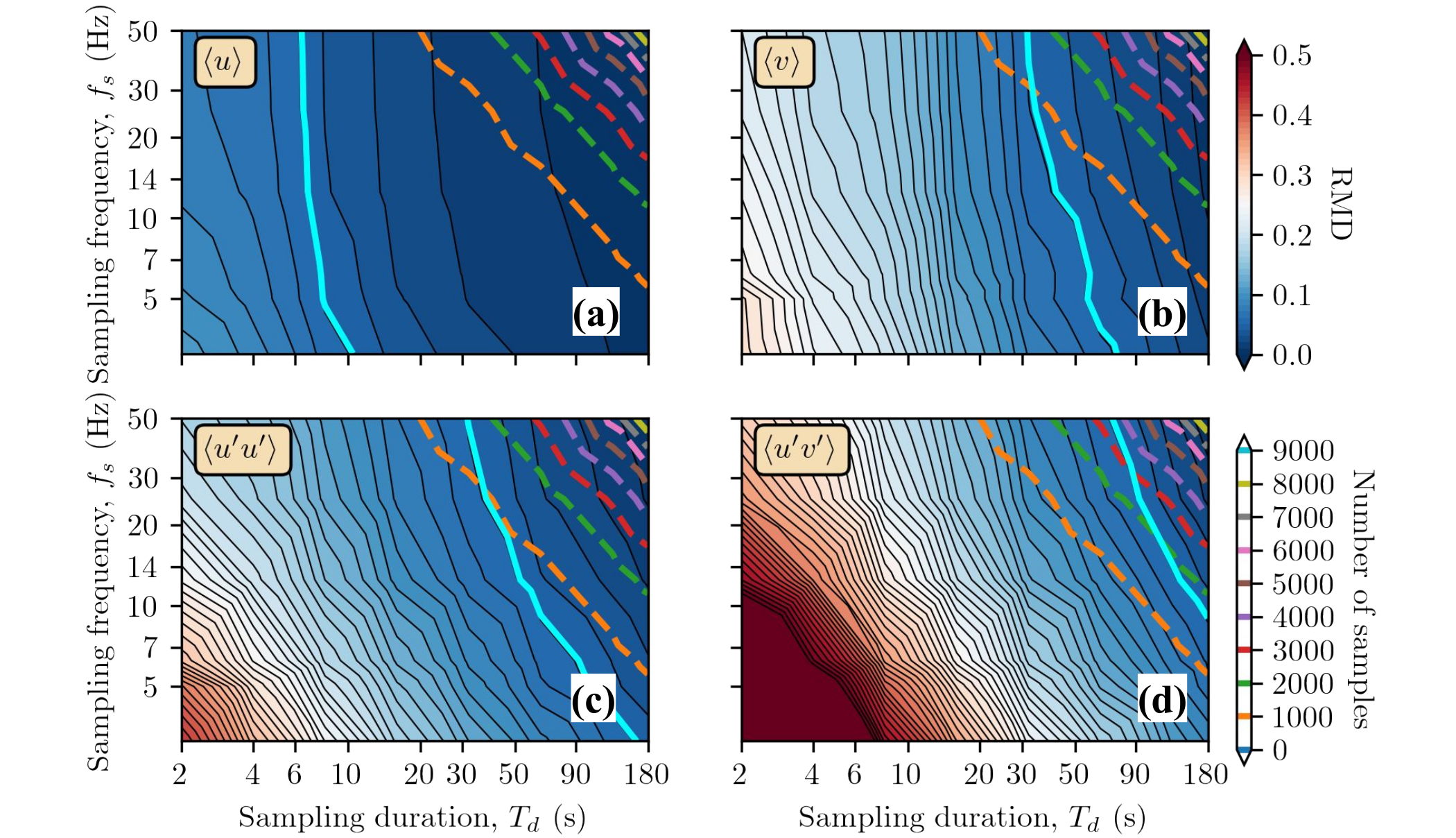}
    \caption{Dependence of statistics convergence on the sampling frequency $f_s$ and sampling duration $T_d$. (a) streamwise mean velocity $\langle u \rangle$, (b) spanwise mean velocity $\langle v \rangle$, (c) Reynolds stress component $\langle u^\prime u^\prime \rangle$, (d) Reynolds stress component $\langle u^\prime v^\prime \rangle$. $\langle v^\prime v^\prime \rangle$ behavior is similar to the $\langle u^\prime u^\prime \rangle$ and is shown in Fig. S6. The relative mean difference (RMD) is defined in Eq. \ref{eq:RMD}. PIV measurements for case 4 are used.}
    \label{fig:statistics_convergence}
\end{figure}

Fig. \ref{fig:statistics_convergence} illustrates the RMD contours for statistical convergence as functions of $f_s$ and $T_d$. Additionally, the solid cyan lines denote the 5\% RMD thresholds. For the mean velocity components $\langle u \rangle$ and $\langle v \rangle$, Fig. \ref{fig:statistics_convergence} shows that statistical convergence improves with increases in both $T_d$ and $f_s$. Specifically, the color maps in the top row indicate that the RMD decreases significantly as both $T_d$ and $f_s$ increase. The dark blue regions, which denote smaller RMD values, become more prominent with higher sampling frequencies and longer durations, signifying better convergence. Compared to the first-order statistics, the second-order statistics of Reynolds stress, $\langle u^\prime u^\prime \rangle$ and $\langle u^\prime v^\prime \rangle$, demonstrate a slower statistical convergence. This is illustrated by the extensive dark red regions in the bottom plot of Fig. \ref{fig:statistics_convergence}; the threshold line is also shifted towards higher samples. This is particularly evident for the Reynolds shear stress $\langle u^\prime v^\prime \rangle$. Achieving the same degree of convergence for higher-order moments requires larger sample sizes, nearly 5x that required for first-order statistics. The behavior of the $\langle v^\prime v^\prime \rangle$ statistics, as shown in Fig. S6, also yields a similar convergence pattern to that of the $\langle u^\prime u^\prime \rangle$ statistics.

Also plotted in Fig. \ref{fig:statistics_convergence} is the number of total samples (dashed lines). The total sample number can directly correlate to the computational time needed for the PIV algorithm processing. Notably, the contour lines of the total samples are not parallel to those of the RMD but instead display a distinct angle. This deviation is more pronounced for the lower-order momentum, as shown in Figs. \ref{fig:statistics_convergence}a and b. This finding suggests that for an equivalent number of samples and computational processing time, a strategy involving longer sampling durations and lower sampling frequencies yields more statistically converged results compared to shorter durations with higher frequencies. This approach is particularly effective when the primary objective is to obtain lower-order flow statistics. Another advantage of using a lower frequency and longer sampling period is the reduced demand for PIV hardware, specifically concerning the laser frequency and camera requirements. High-resolution, high-speed cameras often necessitate additional frame grabbers and substantial RAM for frame storage (i.e., in this study, each PIV run requires at least 216 GB RAM). For higher-order statistics, such a sampling strategy becomes less effective, as evidenced by the smaller angle between the RMD contour and total sample contour lines. Given that analyzing the temporal dynamics of turbulent flow is also a key objective of this study, a sampling frequency of 50 Hz and a sampling duration of 180 seconds is utilized in subsequent analyses. These parameters have been demonstrated to be more than sufficient for robust flow statistics.

\subsection{Mean flow statistics and Reynolds stress}
\begin{figure}[H]
    \centering
    \includegraphics[width=1.0\textwidth]{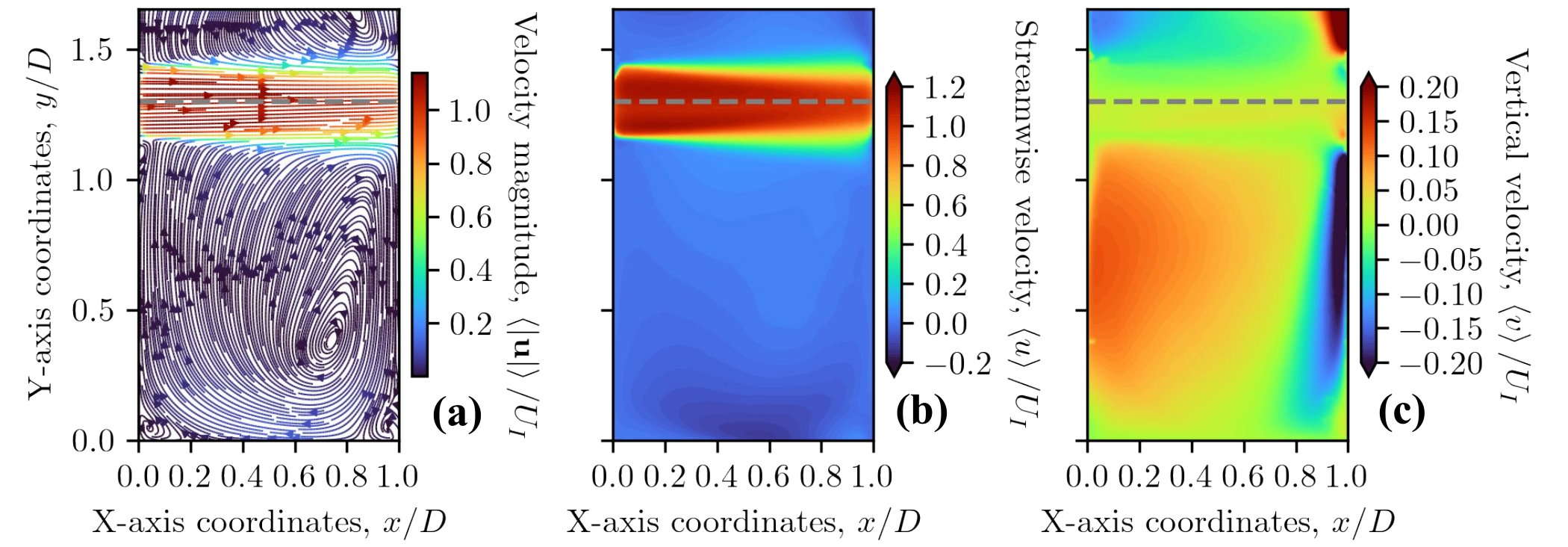}
    \caption{Pseudo 2D flow streamlines and mean flow velocities. (a) pseudo 2D streamlines are colored by the normalized velocity magnitude $\left |\langle\mathbf{u} \right \rangle|/ U_I$, (b) mean streamwise velocity $\langle u \rangle/ U_I$, (c) mean vertical velocity $\langle v \rangle/ U_I$. $U_I$ is the inlet bulk velocity. Inlet flow direction is from left to right. PIV measurements for case 4 are used.}
    \label{fig:mean_flow}
\end{figure}

Fig. \ref{fig:mean_flow} illustrate the general flow patterns in the bench-scale HS. The overall flow structure can be classified as a jet-driven cavity flow, with the formation of vortices and the presence of recirculation zones. As illustrated by the pseudo flow streamlines in Fig. \ref{fig:mean_flow}a, a large recirculation zone is observed in the lower half of the domain, extending from $y/D = 0.15$ to $y/D = 1.0$. This recirculation zone is characterized by circular streamlines, suggesting a significant vortex structure. Additionally, two smaller circulation regions are also observed at the two lower corners of the system. Above the main circulation zone, the streamlines become more parallel, indicating a transition to a more uniform flow. This behavior is consistent with the formation of a shear layer between the recirculating flow and the incoming free stream.

Fig. \ref{fig:mean_flow}b shows the distribution of the mean streamwise velocity component, normalized by the inlet bulk velocity $U_I$. The high-velocity region concentrated near the top of the domain, extending from $y/D = 1.14$ to $y/D = 1.43$ is the inlet jet region. The radial expansion of the jet is observed. However, due to the limited extent, the development of the inlet jet is constrained and the classic conical shape and self-similarity are not yet formed (typically require development distance $>$ 20$D_I$) \citep{Pope2000}. Below this jet, the velocity magnitude decreases sharply, transitioning to lower values in the recirculation zone. The sharp velocity gradient at the interface between the high-velocity jet and the recirculating flow highlights the presence of strong shear, which contributes to the observed flow instability and turbulence, as shown in the later section. Fig. \ref{fig:mean_flow}c illustrates the normalized vertical velocity component across the domain, where positive and negative values represent upward and downward flows, respectively. In the lower recirculation zone, the vertical velocity is predominantly negative, reaching values as low as $-0.2 U_I$. This downward flow corresponds with the vortex structure observed in the streamline plot and strongly indicates the impingement of the inlet jet. Above this zone, the vertical velocity transitions to positive values, reflecting upward movement within the high-velocity jet region. This transition is characterized by a horizontal shear layer, where the mean vertical velocity approaches zero. Combined with Figs \ref{fig:mean_flow}a and c, these observations suggest that the expansion and impingement of the inlet jet are primary drivers of the internal flow hydrodynamics in the HS system. The flow circulation, in turn, provides feedback to the inlet jet, modifying the shear layer and further influencing the flow dynamics \citep{Kang2009}. 

\begin{figure}[H]
    \centering
    \includegraphics[width=1.0\textwidth]{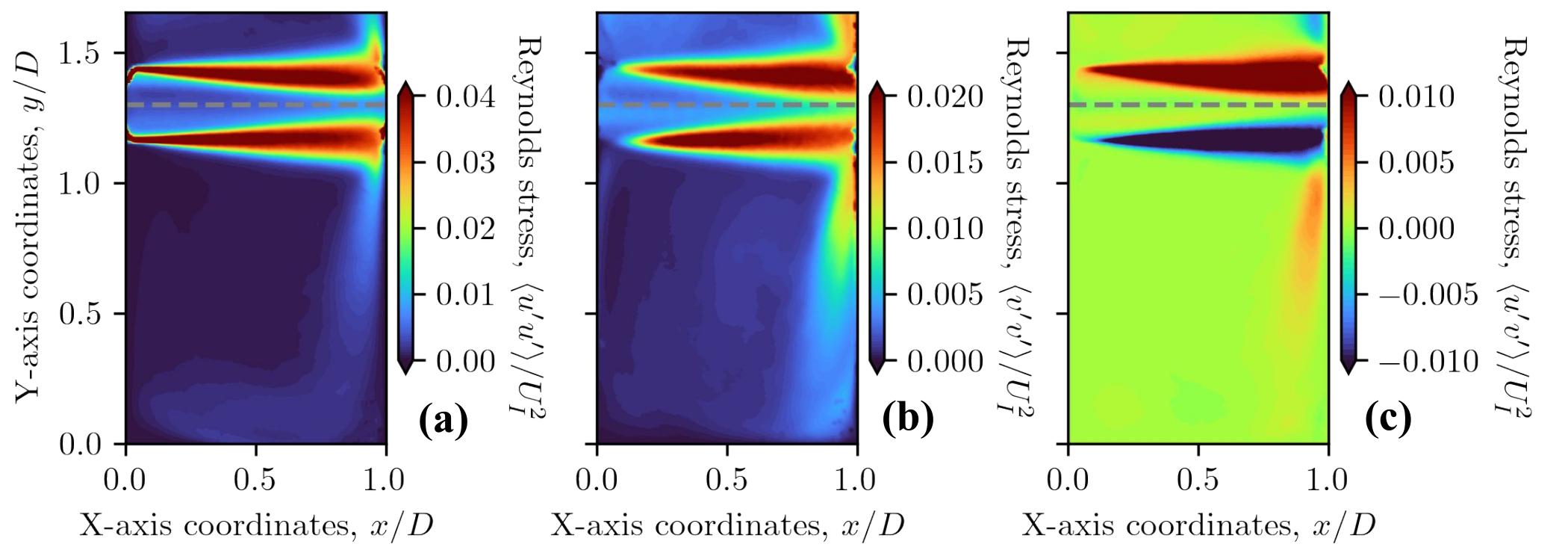}
    \caption{Second-order flow statistics of Reynolds stress. (a) normalized Reynolds stress component $\langle u^\prime u^\prime \rangle/ U_I^2$, (b) normalized Reynolds stress component $\langle v^\prime v^\prime \rangle/ U_I^2$, and (c) normalized Reynolds stress component $\langle u^\prime v^\prime \rangle/ U_I^2$. $U_I$ is the inlet bulk velocity. Inlet flow direction is from left to right. PIV measurements for case 4 are used.}
    \label{fig:RSS}
\end{figure}

Fig. \ref{fig:RSS} presents the spatial distribution of normalized Reynolds stresses within the flow field, which are critical indicators of turbulence intensity and anisotropy. In Fig. \ref{fig:RSS}a, the distribution of the Reynolds stress component $\langle u^\prime u^\prime \rangle$ is depicted. The contour plot reveals that the highest values of $\langle u^\prime u^\prime \rangle$ are concentrated near the top of the domain, specifically around $y/D = 1.14$. This region aligns with the invert line of the inlet pipe, where a high-velocity jet is observed in the mean velocity plots, indicating significant streamwise velocity fluctuations. The intensity of these fluctuations diminishes both downstream and away from the jet, transitioning from red to blue regions on the plot. The presence of strong velocity gradients and high shear in this region likely contributes to the elevated levels of $\langle u^\prime u^\prime \rangle$. Fig. \ref{fig:RSS}b illustrates the spatial distribution of the Reynolds stress component $\langle v^\prime v^\prime \rangle$. Similar to $\langle u^\prime u^\prime \rangle$, the highest values of $\langle v^\prime v^\prime \rangle$ are observed near the top of the domain. This region of high vertical velocity fluctuations overlaps with the high-velocity jet, suggesting that both streamwise and vertical velocity components are highly turbulent in this area. The contour plot displays a symmetric pattern around the centerline, with the magnitude of $\langle v^\prime v^\prime \rangle$ decreasing towards the recirculation zone. This symmetry and distribution suggest that the vertical fluctuations are also significant contributors to the overall turbulence within the flow. Fig. \ref{fig:RSS}c shows the Reynolds shear stress component $\langle u^\prime v^\prime \rangle$, which represents the covariance between the streamwise and vertical velocity components. The larger negative values of $\langle u^\prime v^\prime \rangle$ indicate that the fluctuations in $u^\prime$ and $v^\prime$ are negatively correlated, signifying turbulence anisotropy. This pattern suggests that downward-moving eddies ($v^\prime < 0$) in the shear layer transport higher streamwise momentum ($u^\prime > 0$) to lower regions. Regions displaying higher values of $\langle u^\prime v^\prime \rangle$ are closely aligned with the shear layer, highlighting the role of turbulent shear stresses in facilitating momentum transfer across the interface between the jet and the recirculation zone.

\subsection{Influence of Reynolds number on flow statistics}
\begin{figure}[H]
    \centering
    \includegraphics[width=1.0\textwidth]{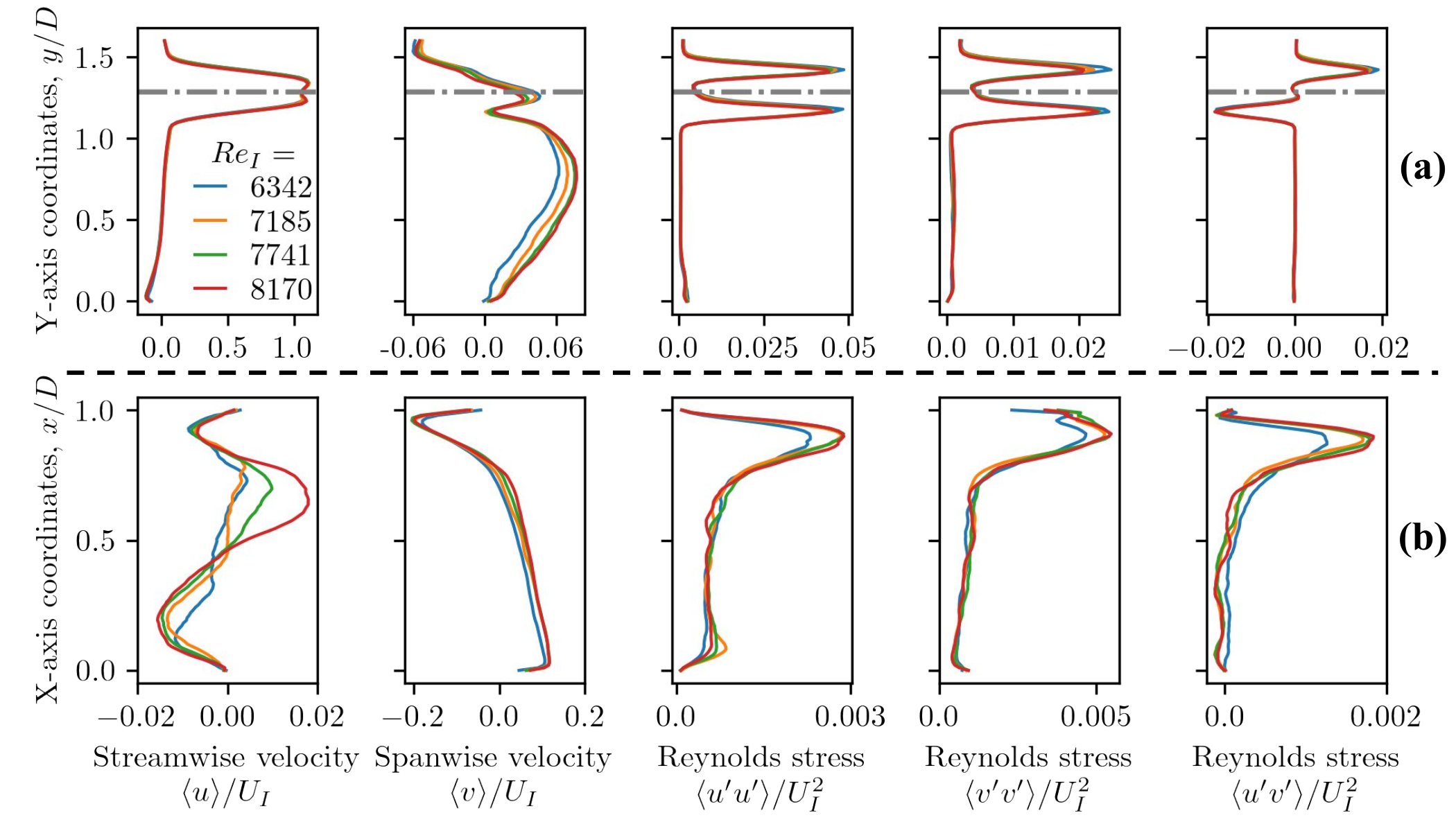}
    \caption{Dependence of mean flow statistics and Reynolds stress on Reynolds number. (a) profile along the $y$-axis coordinates at the line $x/D = 0.5$, (b) profile along the $x$-axis coordinates at the line $y/D = 0.5$.}
    \label{fig:Re_dependance}
\end{figure}

Dimensional analysis and dynamic similarity indicate that for internal flows, the Reynolds number is the primary parameter influencing the nondimensional flow solution. Fig. \ref{fig:Re_dependance}a evaluates mean flow statistics and Reynolds stresses along the $y$-axis at the line $x/D = 0.5$ for the four cases with inlet flow Reynolds numbers of 6342, 7185, 7741, and 8170. Both mean flow and Reynolds stress exhibit limited dependence on the Reynolds number, with statistics from different Reynolds numbers nearly overlapping. However, the spanwise velocity demonstrates some sensitivity to the Reynolds number changes. Specifically, as the Reynolds number increases, the upward flow in the main circulation region also increases. Nonetheless, this observed dependence is less pronounced than that reported in a full-scale HS system by \citet{Howard2010thesis}.

A comparison of flow statistics and Reynolds stresses along the $x$-axis at the line $y/D = 0.5$ also reveals a disparity in their dependence on the Reynolds number, as illustrated in Fig. \ref{fig:Re_dependance}b. In particular, the streamwise velocity profile at this location shows significant sensitivity to changes in the Reynolds number. As the Reynolds number increases, a notable increase in streamwise velocity is observed at $x/D = 0.7$. Pseudo 2D flow streamlines, depicted in Fig. S7 (online Supplemental Materials), show that the line $y/D = 0.5$ intersects the main circulation eye. With increasing Reynolds number, the circulation eye tends to shift towards the lower right region of the domain. Meanwhile, other statistics remain similar across different Reynolds numbers, as shown in Fig. S8 (online Supplemental Materials). Overall, these results, despite some deviations, generally align with the phenomenon of Reynolds number invariance, which is often observed in turbulent flows at higher Reynolds numbers \citep{Heller2017, Li2022SOR}.

\subsection{Instantaneous flow structure and spectral characteristics}
\begin{figure}[H]
    \centering
    \includegraphics[width=1.0\textwidth]{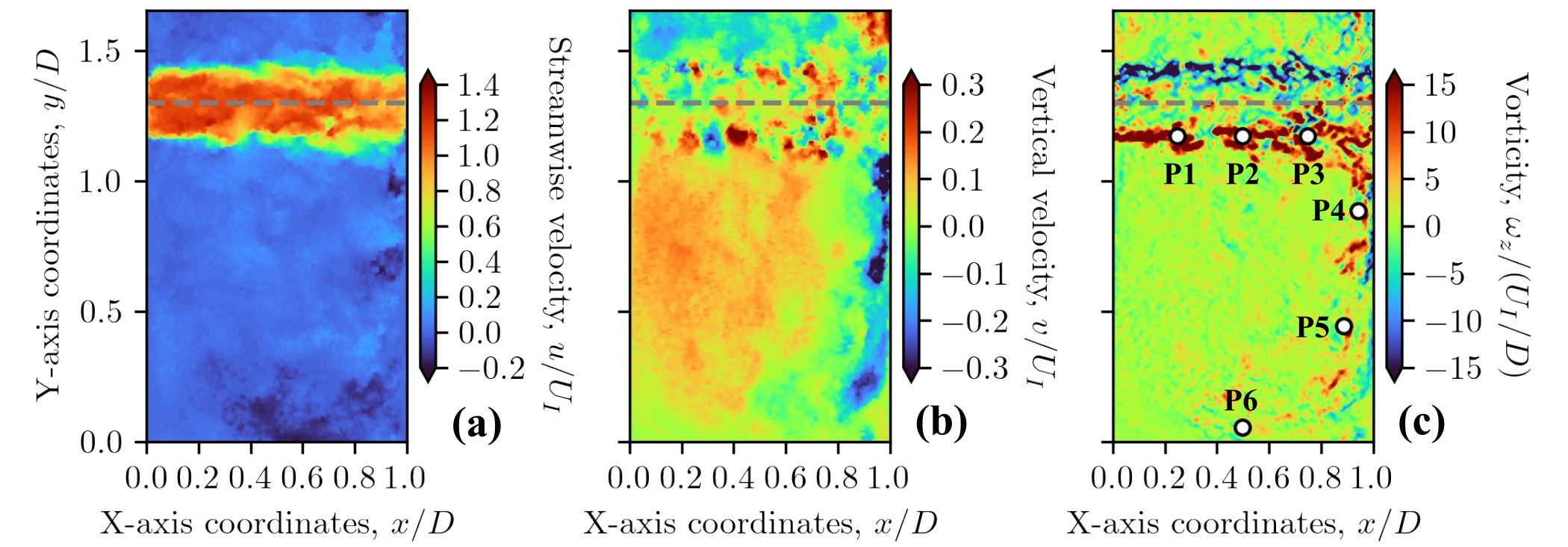}
    \caption{Instantaneous flow velocities and vortices contours. (a) streamwise velocity $u/U_I$, (b) vertical velocity $u/U_I$, and (c) spanwise vorticity $\omega_z/(U_I/D)$. Case 4 is shown as an example. Points in plot (c) show the prob locations. An animation is shown in Video S1 in online Supplemental Materials.}
    \label{fig:insta_flow}
\end{figure}

Fig. \ref{fig:insta_flow} and Video S1 (online Supplemental Materials) illustrate the instantaneous flow velocities and vorticity contours in the HS model. Smaller vortices characterized by high rotational velocities are observed in the shear layers of the inlet jet, as indicated by elevated vorticity values. As the shear layer develops and expands, these vortices grow in size and become entrained with the upward flow from the main circulation. At the outlet, the radial expansion of the inlet jet surpasses the outlet diameter, causing the outer regions of the jet to impinge upon the lips of the outlet. This interaction redirects the streamwise momentum downward into vertical motion, forming a high-speed jet along the outlet wall. This downward jet continues to develop along the sidewall, eventually deflecting away from the wall and spreading lengthwise due to the mean flow recirculation in the lower right region of the model. This impinging jet plays a crucial role in feeding momentum to the main circulation in the bulk region of the HS model. As the wall jet progresses, vortices begin to hover away from the walls and interact with the slower flow in the bulk region, gradually losing their strength and dissipating their energy. The vortices along the wall jet and the bottom wall exhibit much larger structures with reduced swirling speeds, indicating a tendency toward laminarization. The upflow in the circulation region displays a significantly larger spatial extent compared to the downward jet, with decreased vortical structures.

\begin{figure}[H]
    \centering
    \includegraphics[width=1.0\textwidth]{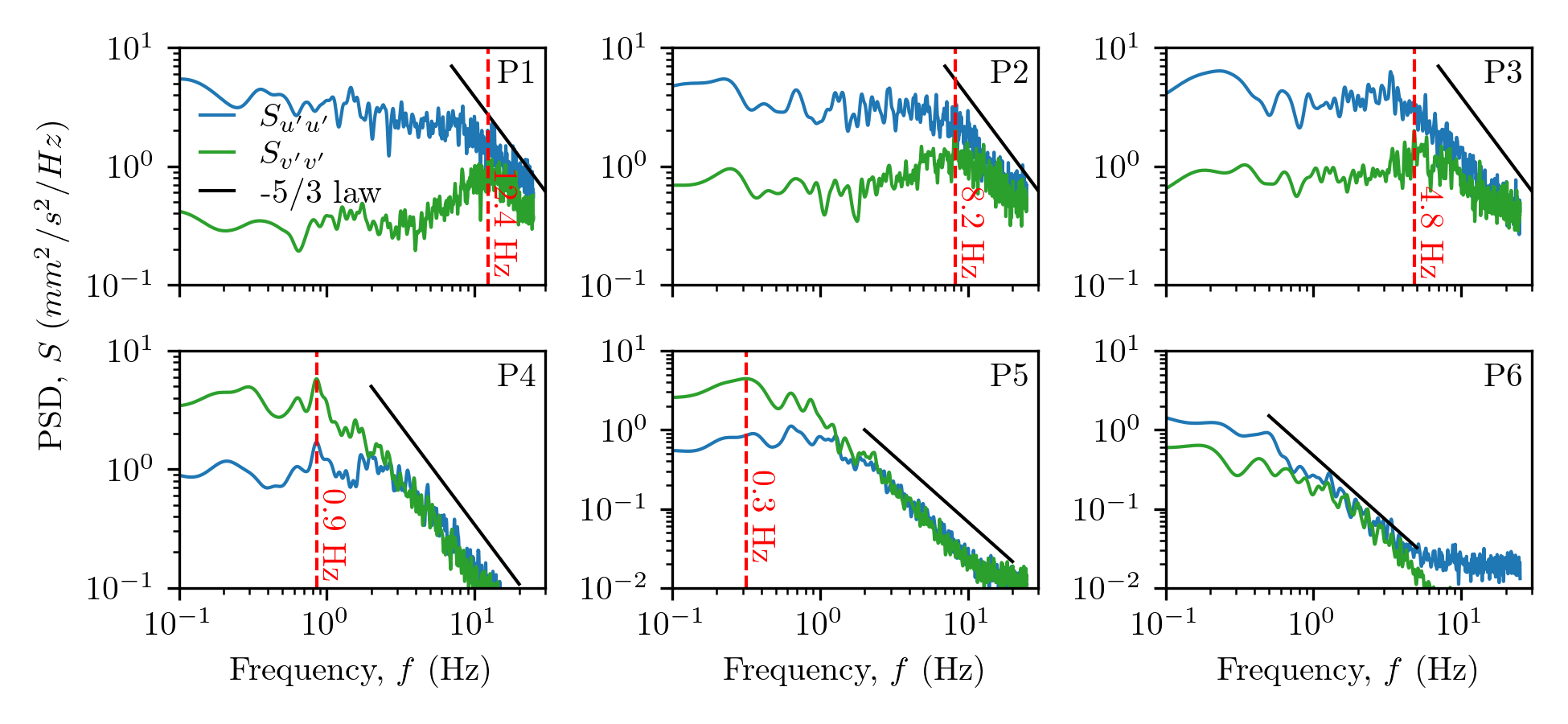}
    \caption{Power spectral densities (PSD) of streamwise velocity fluctuation $u^\prime$ and vertical velocity fluctuation $v^\prime$. The probe locations are shown in Fig. \ref{fig:insta_flow}c. The red dash line outline the peak frequency for vertical velocity component $v^\prime$. The solid black line indicates the Kolmogorov's $E(f) \propto f^{-5/3}$ power law \citep{Pope2000}.}
    \label{fig:spectral_analysis}
\end{figure}

Video S1 (online Supplemental Materials) illustrates the unsteadiness and intermittency of the turbulent flow within the HS model. Particularly at the impinging region near the outlet lips, the downward jet exhibits strong fluctuations with periodic vortex bursts. Combined with the vorticity distribution shown in Fig. \ref{fig:insta_flow}c, a clear pathway of vortices from the inlet shear layer to the HS bulk volume can be traced. To quantify these dynamics, six probing points are configured along the vorticity transport pathway, as shown in Fig. \ref{fig:spectral_analysis}c. The power spectral densities (PSD) of streamwise velocity fluctuations $u^\prime$ and vertical velocity fluctuations $v^\prime$ at these respective locations are presented in Fig. \ref{fig:spectral_analysis}. The red dashed line highlights the peak frequency for the vertical velocity component $v^\prime$. Additionally, the solid black line indicates adherence to Kolmogorov's $E(f) \propto f^{-5/3}$ power law, characteristic of the inertial subrange in turbulent flows \citep{Kolmogorov1991, Pope2000}.

At probe locations P1, P2, and P3 (top row), the energy of streamwise velocity fluctuations $S_{u'u'}$ is spread across the spectrum, showing higher energy content at lower frequencies, which gradually decreases with increasing frequency. In contrast, the vertical velocity fluctuations $S_{v'v'}$ exhibit a different pattern, with a distinct energy peak at higher frequencies, an indicator of the presence of coherent turbulent structures. Along the development of the shear layer, the energy peak gradually shifts towards lower frequencies. This observation is in agreement with the findings from \citep{Mendez2019}, where the vortex structure is observed to stretch and enlarge as the shear layer develops. Irrespective of velocity components, the -5/3 power law is overlaid on these plots at higher frequencies, indicating a region of the inertial subrange where the energy cascade follows this scaling law. The presence of this scaling law suggests that the turbulence at these locations is well-developed and exhibits the typical characteristics of turbulent energy transfer. The observed inertial range is truncated by the PIV measurement temporal resolution of 50 Hz, but it is expected to extend further into higher frequencies.

Probe locations P4, P5, and P6 (bottom row) focus on the wall jet development region. In these regions, the energy levels between $S_{u'u'}$ and $S_{v'v'}$ reverse due to wall deflection. A distinct energy peak at \SI{0.9}{Hz} is observed at Probe P4, likely associated with the coherent vortex bursting behavior created by the intermittent passage of the primary vortices as seen in Video S1 (online Supplemental Material). The peak frequency is found to be case-specific and varies across different Reynolds numbers, as shown from Fig. S9. The inertial subrange (-5/3 power law) is also observed at these probe locations. The streamwise velocity fluctuations at P6 exhibit an elevated spectrum after the inertial subrange, rather than a rapid drop-off due to viscous dissipation. Upon interrogation of the measured data, this spectrum flattening behavior is potentially caused by measurement noise at higher frequencies (e.g., laser sheet fluctuation).

\subsection{Modal analysis with POD and SPOD}
\begin{figure}[H]
    \centering
    \includegraphics[width=0.95\textwidth]{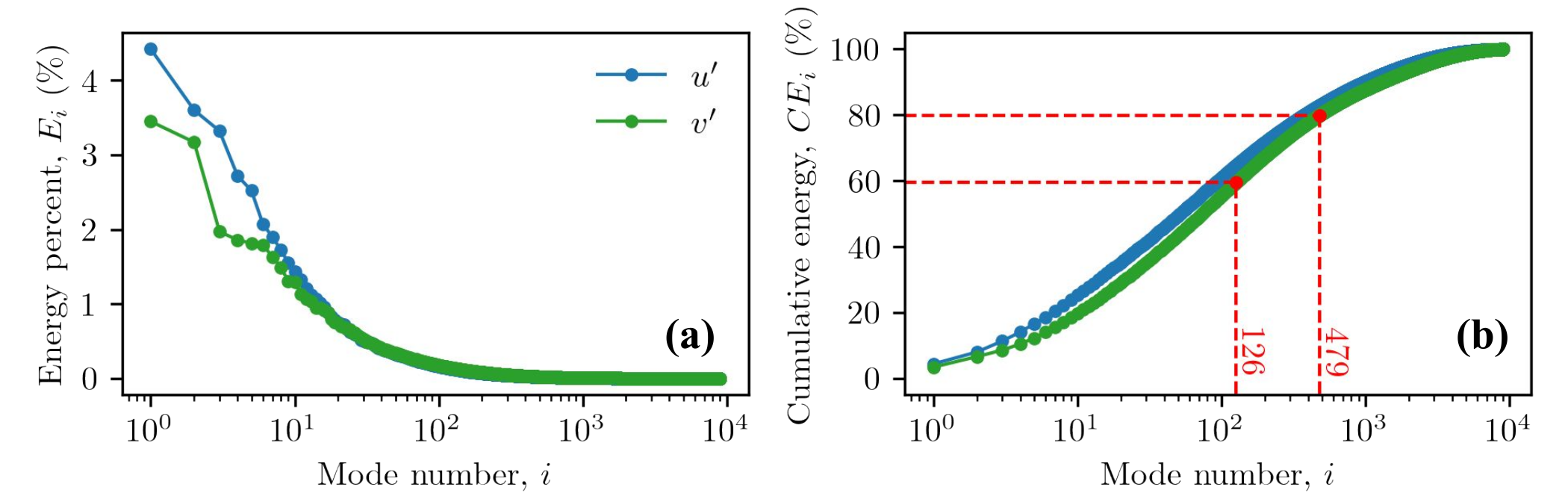}
    \caption{POD decomposition of streamwise and vertical velocity fluctuation fields $u^{\prime}$ and $v^{\prime}$. (a) energy percent $E_i$ as defined in Eq. \ref{eq:POD_energy}, (b) cumulative energy ${CE}_{i}$ as defined in Eq. \ref{eq:POD_cum_energy}.}
    \label{fig:POD_energy}
\end{figure}

Two mode decomposition methods, POD and SPOD, are applied to the instantaneous streamwise and vertical velocity fluctuation fields $u^{\prime}$ and $v^{\prime}$, i.e., the mean flow is removed. Fig. \ref{fig:POD_energy} illustrates the energy percent $E_{i}$ and the cumulative energy $CE_{i}$, as defined in Eqs. \ref{eq:POD_energy}-\ref{eq:POD_cum_energy}.

\begin{equation}
    E_i = \frac{\lambda_i}{\sum_{p=1}^{n} \lambda_p}, \label{eq:POD_energy}
\end{equation}

\begin{equation}
    {CE}_i = \sum_{q=1}^{i} E_q. \label{eq:POD_cum_energy}
\end{equation}
In these equations, where $\lambda_{i}$ is the eigenvalue associated with mode $i$, and $n$ is the total number of modes.

Distinct from fluid flows that exhibit dominant lower-order dynamics, such as laminar flow over a cylinder, POD decomposition reveals that turbulent flow inherently possesses higher dimensions. Unlike being concentrated within a few dominant frequencies, the kinetic energy in turbulent flow is distributed across a wider range of scales and frequencies. This complexity is highlighted by the requirement of approximately 479 modes to capture 80\% of the kinetic energy, as demonstrated in Fig. \ref{fig:POD_energy}b. The energy distribution for $u^\prime$ and $v^\prime$ shows similar behavior. In both cases, there exist paired modes: each pair consisting of modes with similar eigenvalues but phase-shifted structures. Previous studies \citep{FazleHussain1986, Taira2017, Towne2018} have shown that these pairing modes are crucial for representing certain turbulent structures, such as traveling waves, which could not be well-represented by individual modes. This phenomenon is illustrated by the spatial distribution of the first 100 POD modes in Fig. S10 (online Supplemental Material). All the $u^\prime$ modes in the range of $i$ from 2 to 30 tend to be related to traveling turbulence structures and, therefore, arise in pairs. Video S2 demonstrates how the temporal dynamics of these paired modes are closely interrelated, and together, they encapsulate the complete oscillatory behavior of the shear layer.

\begin{figure}[H]
    \centering
    \includegraphics[width=1.0\textwidth]{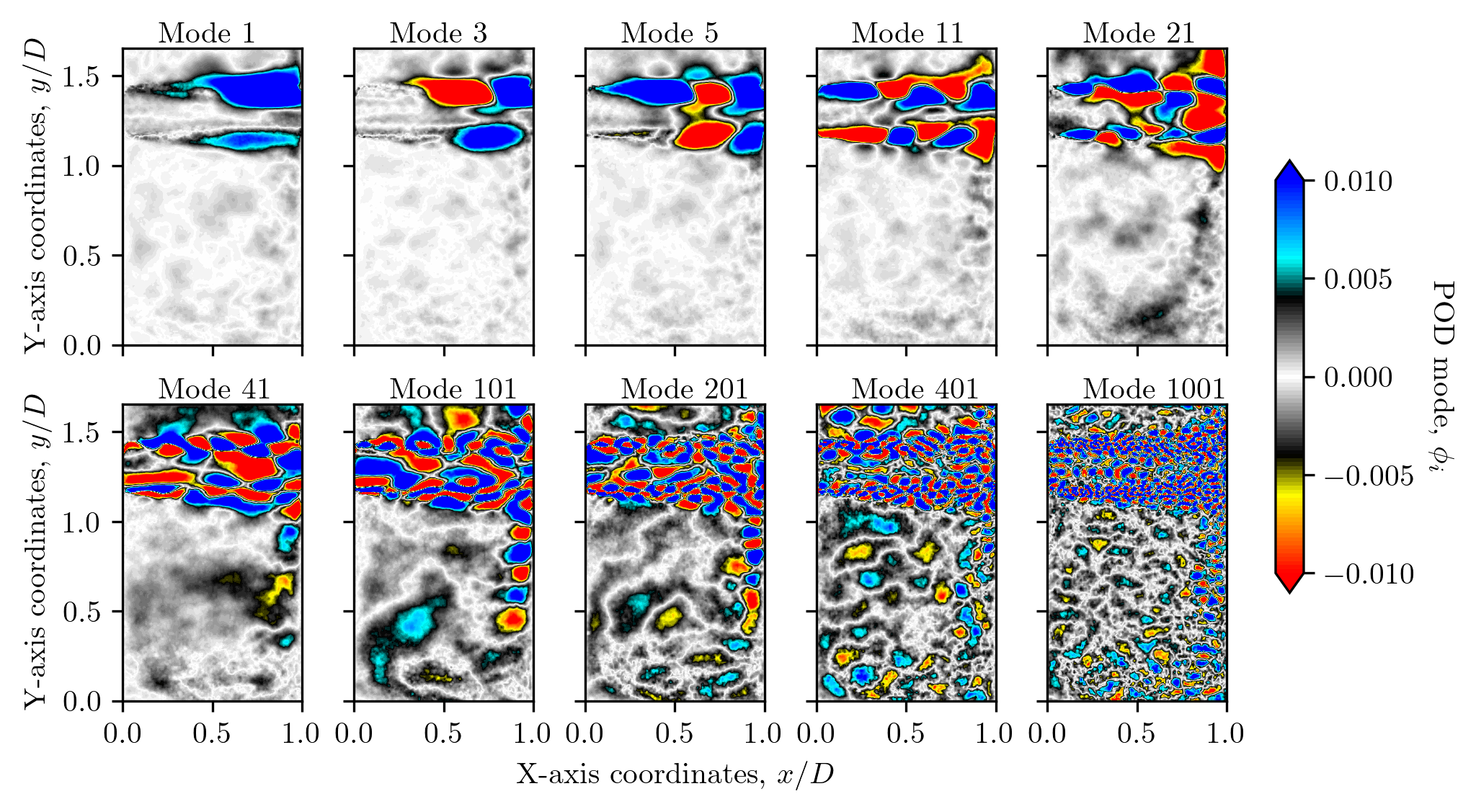}
    \caption{POD mode of streamwise velocity fluctuation. The odd mode numbers are selected for illustration. The first 100 POD modes in Fig. S10 in the online Supplemental Material.}
    \label{fig:POD_mode_u}
\end{figure}

\begin{figure}[H]
    \centering
    \includegraphics[width=1.0\textwidth]{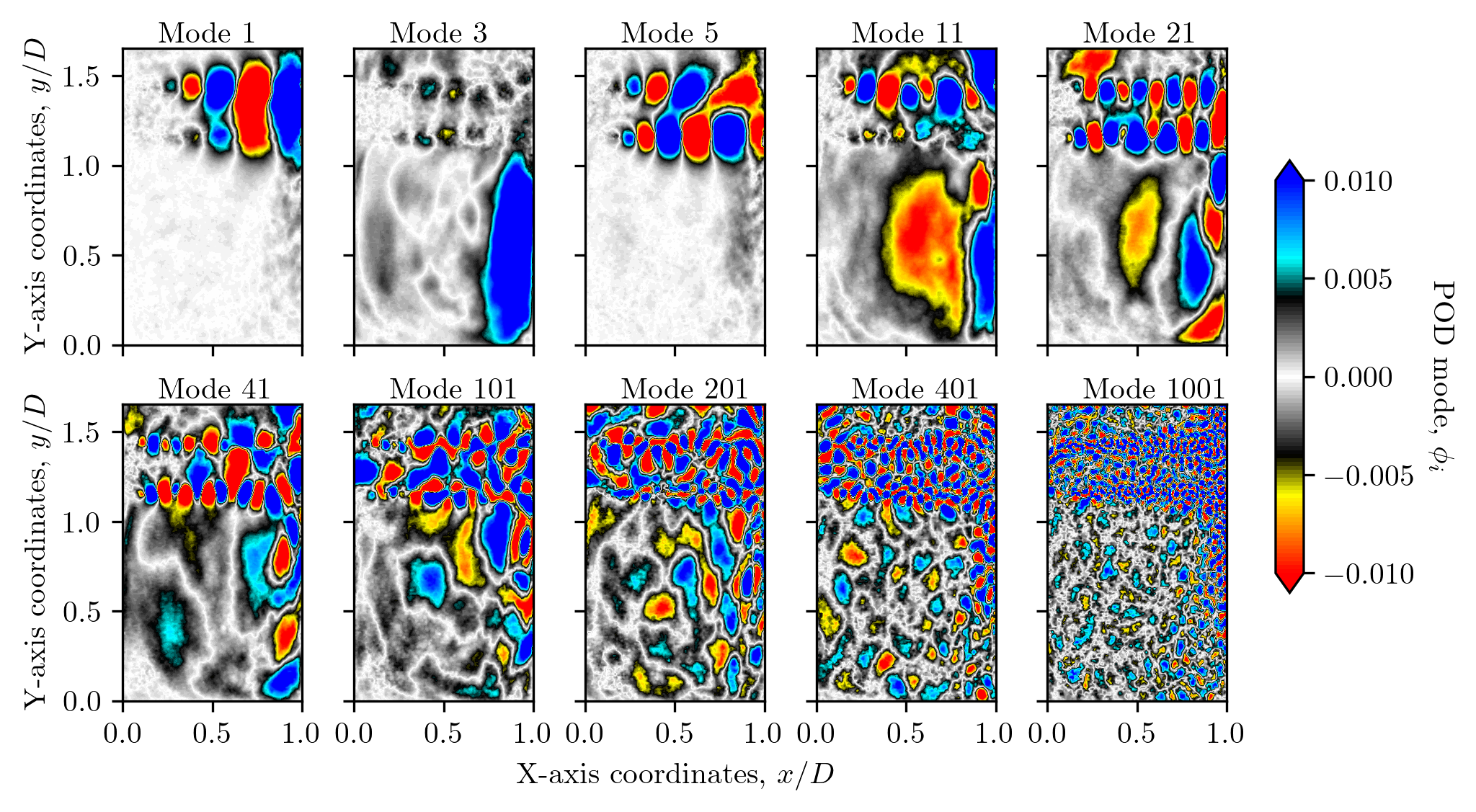}
    \caption{POD mode of vertical velocity fluctuation. The odd mode numbers are selected for illustration. The first 100 POD modes in Fig. S11 in the online Supplemental Material.}
    \label{fig:POD_mode_v}
\end{figure}

Figs. \ref{fig:POD_mode_u} and \ref{fig:POD_mode_v} illustrate the spatial patterns of the representative POD modes for $u^{\prime}$ and $v^{\prime}$. In lower-order modes, such as Mode 1, Mode 3, and Mode 5, the structures are large and coherent, representing the most energetic and dominant features of the flow. These modes capture the primary dynamics of the flow. For instance, Mode 1 in $u^{\prime}$ and Mode 3 in $v^{\prime}$ depict the large-scale oscillations in the inlet jet and impinged wall jet. Modes 3, 5, 11, and 21 of $u^{\prime}$ and Modes 1, 5, 11, and 21 of $v^{\prime}$ capture energetic vortex streets. As the mode number increases, the spatial structures become progressively smaller and more fragmented, indicative of finer-scale turbulence. For example, higher-order modes like Mode 401 and Mode 1001 capture small-scale eddies and less coherent structures, which, although contributing to the overall turbulence, contain minimal energy (less than 0.05\%). The transition from large, coherent structures in low-order modes to small, fragmented structures in high-order modes highlights the multi-scale nature of turbulence, making the isolation of different coherent turbulent structures challenging. Indeed, the higher-order modes in Figs. \ref{fig:POD_mode_u} and \ref{fig:POD_mode_v} tend to reveal a mixture of vortex streets from the inlet jet and vortices in the main circulation. 

\begin{figure}[H]
    \centering
    \includegraphics[width=0.95\textwidth]{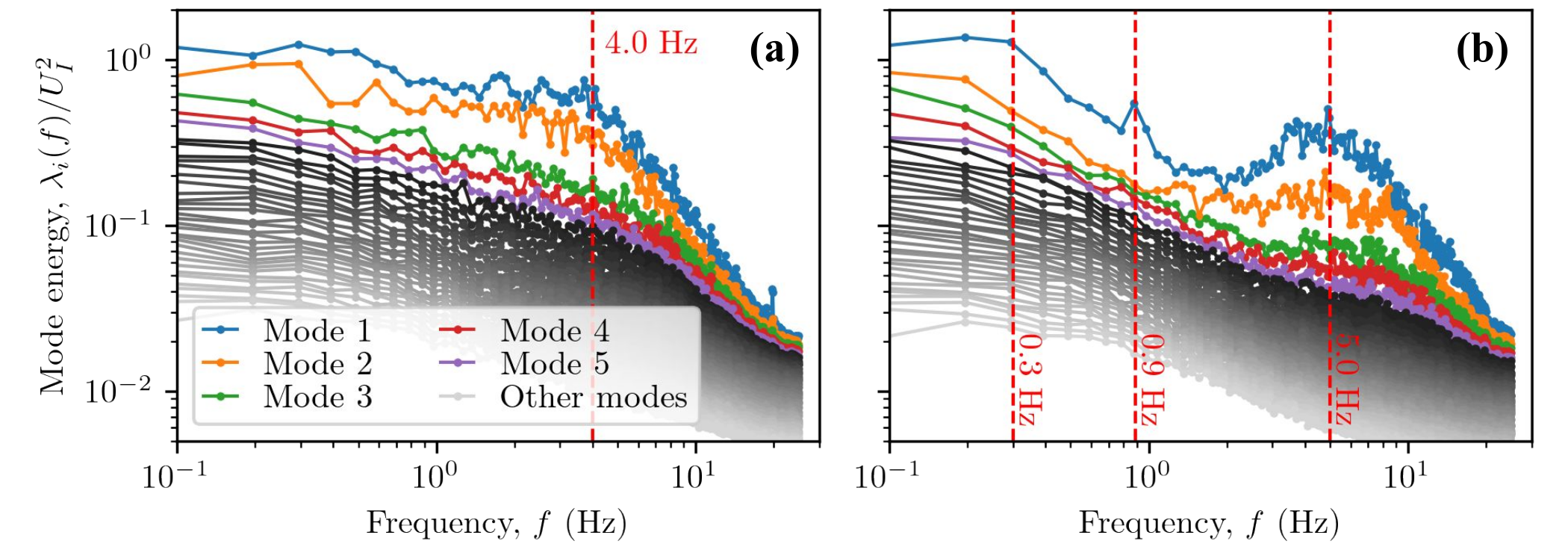} 
    \caption{SPOD decomposition of velocity fluctuation fields. (a) streamwise velocity $u^{\prime}$ (b) vertical velocity $v^{\prime}$. $\lambda_i(f)$ is the energy (or eigenvalue) of $i$-th mode at $f$ frequency. The mode energy is normalized by the inlet bulk velocity $U_I$. The first 5 leading modes are illustrated and higher-order modes are shown as black to gray lines.}
    \label{fig:SPOD_energy}
\end{figure}

In contrast to POD, SPOD considers frequency information and facilitates the decomposition of the flow field into modes that are coherent in both space and time. Fig. \ref{fig:SPOD_energy} shows the SPOD decomposed energy (eigenvalue) distribution across different modes and frequencies. The lower-order modes, Mode 1 and Mode 2, exhibit higher energy content compared to the higher-order modes, Mode 3, Mode 4, and Mode 5. The energy distribution pattern of the lower-order modes depends on the frequency and the specific velocity field. However, the higher-order modes display a broader energy distribution across the entire frequency range. It is also noted that lower modes (e.g., Mode 1, Mode 2) do not necessarily always yield significantly higher energy, especially at higher frequencies. This suggests that the dynamics in these regions can be less coherent and do not contain energy-dominant flow structures. In the streamwise velocity fluctuation $u^{\prime}$, larger energy separations between leading modes and higher-order modes are generally observed from lower frequency up to $f$ = 5 Hz, with the highest separation occurring approximately at $f$ = 4 Hz. For the vertical velocity fluctuation $v^{\prime}$, energy separations show different behavior with three energy peaks (\quotes{energy humps}) at frequencies 0.3, 0.9, and 5 Hz. Note that in the previous spectral analysis \ref{fig:spectral_analysis}, the PSD at probes P4 and P5 also yielded distinct peak frequencies of 0.9 and 0.3 Hz.

\begin{figure}[H]
    \centering
    \includegraphics[width=1.0\textwidth]{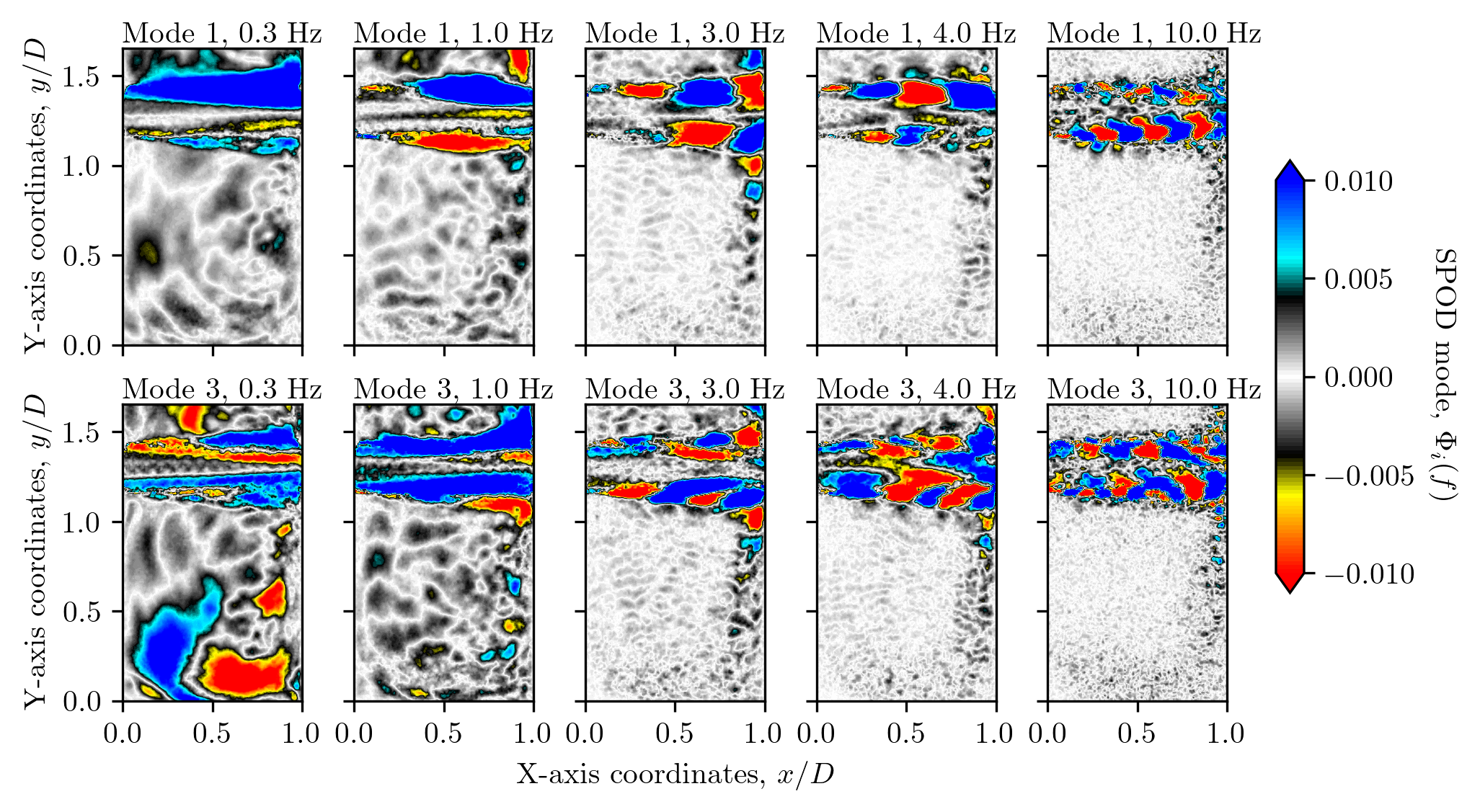}
    \caption{SPOD mode of streamwise velocity fluctuation. Mode 1 and Mode 3 for 5 frequencies are selected for illustration. Mode 1 to 3 for 36 frequencies is provided in Fig. S12 in the online Supplemental Material.}
    \label{fig:SPOD_mode_u}
\end{figure}

\begin{figure}[H]
    \centering
    \includegraphics[width=1.0\textwidth]{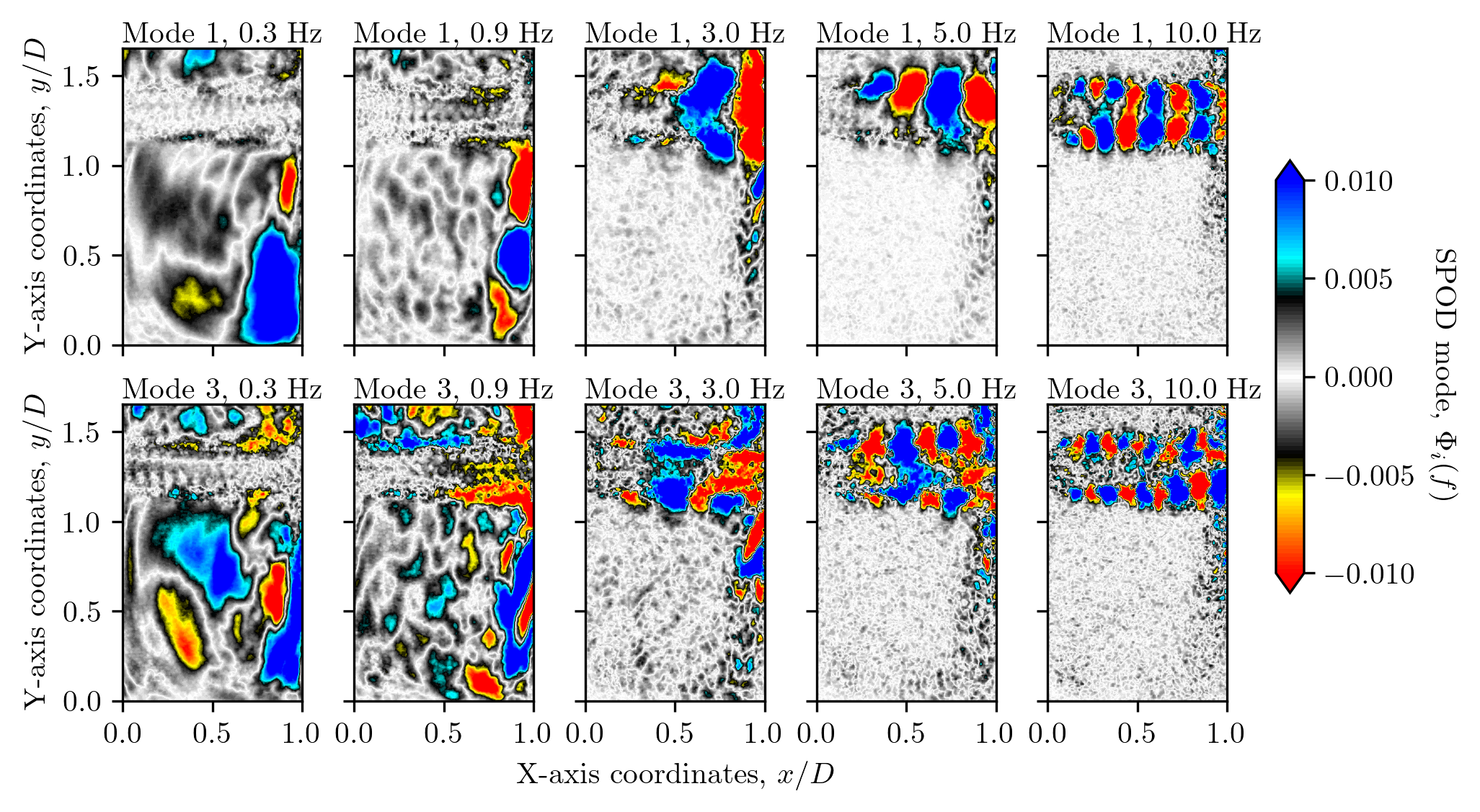}
    \caption{SPOD mode of vertical velocity fluctuation. Mode 1 and Mode 3 for 5 frequencies are selected for illustration. Mode 1 to 3 for 36 frequencies is provided in Fig. S13 in the online Supplemental Material.}
    \label{fig:SPOD_mode_v}
\end{figure}

Figs. \ref{fig:SPOD_mode_u} and \ref{fig:SPOD_mode_v} illustrate the spatial distribution of SPOD Mode 1 and Mode 3 at the representative frequencies identified in Fig. \ref{fig:SPOD_energy}. An expanded visualization of Modes 1, 2, and 3 over the frequency range from 0.2 to 10 Hz is also provided in Figs. S12 and S13 (online Supplemental Materials). In the streamwise velocity fluctuation $u^{\prime}$ of Figs. \ref{fig:SPOD_mode_u}, irrespective of mode rank, SPOD modes with frequencies larger than 3 Hz are all associated with vortex rotation and transport in the inlet shear layer. The leading mode (Mode 1) almost exclusively describes the flow advection and large-scale oscillation of the inlet jet, aligning with the underlying flow physics where the highest streamwise energy is concentrated in the inlet jet. Conversely, SPOD Mode 3 at lower frequencies captures the motion of the cavity flow in the bulk volume of the HS system. These motions have relatively lower energy levels compared to the inlet jet and therefore emerge in the higher-order mode (modes are ranked by their energy level). In the vertical velocity fluctuation $v^{\prime}$ of Fig. \ref{fig:SPOD_mode_v}, the leading modes at lower frequencies are associated with the impinging jet dynamics along the outlet wall, (first two plots in the first row). Specifically, the SPOD Mode 1 at 0.3 Hz and 0.9 Hz identifies the vortex packs that form the bursting motions. These coherent vortex motions are found to be directly responsible for the observed spectral peaks at probes P4 and P5 in Fig. \ref{fig:spectral_analysis}. The spectra at probes P1, P2, and P3 show that the fluid motion in the inlet shear layer has a range of characteristic frequencies from 4.8 to 12.4 Hz, indicating the presence of coherent flow structures. The SPOD analysis also confirms this observation. As shown in Figs. S12 and S13, the SPOD modes in this frequency range also exclusively describe the vortex motions in the inlet jet path.

Compared to POD, SPOD yields a more robust isolation of flow structures. This is particularly illustrated by comparing Fig. \ref{fig:POD_mode_u} and Fig. \ref{fig:POD_mode_v}. In the higher-order POD modes, flow structures of the inlet shear layer and the impinging jet are both present and mixed in space. POD cannot separate these two distinct dynamics due to their similarity in spatial structure and energy level. As revealed in Fig. \ref{fig:SPOD_energy}, the coherent structures associated with the impinging jet (Mode 1 at 0.9 Hz) and inlet shear layer (Mode 1 at 5 Hz) have similar energy levels. As a result, POD fails to clearly distinguish between these dynamics, leading to mixed and less interpretable modes that cannot be assigned to distinct flow phenomena. In contrast, SPOD, by incorporating spectral information, effectively separates these dynamics into distinct modes based on their frequency signatures. This separation is evident in Fig. S12, where the impinging jet and inlet shear layer are isolated into separate modes at their respective dominant frequencies. Because of these enhanced capabilities to identify and isolate coherent structures within complex flow fields, the leading order mode identified by SPOD is shown to be distinctly different from that identified by POD, as demonstrated in Figs. \ref{fig:SPOD_mode_u} and \ref{fig:SPOD_mode_v} and Figs. S12 to S13.

\section{Discussion}

The PIV measurements provide direct visualization of complex turbulent flow in the HS system, including flow separation and jet impinging, as shown in Fig. \ref{fig:insta_flow} and Video S1. These chaotic flow structures directly impact coupled pollutant transport and mixing, and ultimately, the HS treatment performance, such as PM separation. Robust modeling of these dynamics is critical not only for optimizing HS model design but also for assessing the collective behavior of a train of treatment systems at a larger drainage network scale. However, the existing methods are either computationally expensive, which poses challenges for integration into long-term, large-scale simulations such as CFD, or computationally efficient but with lower-fidelity representation of system dynamics, such as those based on lumped dynamical model (e.g., continuous stirred tank reactor, CSTR). An ideal modeling approach would balance lower computational costs with higher-fidelity representations of HS dynamics, potentially leveraging techniques from machine learning or hybrid modeling approaches.

The modal decomposition technique of POD and SPOD, applied in this study, demonstrates a powerful tool for analyzing complex turbulent dynamics and offers another significant benefit of reduced-order modeling (ROM). As shown in Fig. \ref{fig:POD_energy}, 60\% of the (turbulence) kinetic energy is captured with the first 126 modes, and 80\% of kinetic energy is captured with the first 479 modes. The remaining modes only contribute to approximately 20\% of the kinetic energy. Rather than using the full order of the modes, a reduced-order representation of the system dynamics can be reconstructed by only using the leading $r$ number of modes that capture the majority of kinetic energy, as defined in Eq. \ref{eq:POD_reconstruction}. A similar concept is also applicable to SPOD.

\begin{equation}
    \widetilde{\mathbf{X}}^{\prime} = \widetilde{\boldsymbol{\Phi}} \widetilde{\mathbf{A}}. \label{eq:POD_reconstruction}
\end{equation}
In this equation, $\widetilde{(\cdot)}$ indicate the truncated variable using the first $r$ number of modes ($r << n$). $\widetilde{\mathbf{A}} \in \mathbb{R}^{r \times m}$ is the $r$-rank truncated coefficient matrix, $\widetilde{\boldsymbol{\Phi}} \in \mathbb{R}^{n \times r}$ is the $r$-rank truncated eigenvector matrix.

\begin{figure}[H]
    \centering
    \includegraphics[width=1.0\textwidth]{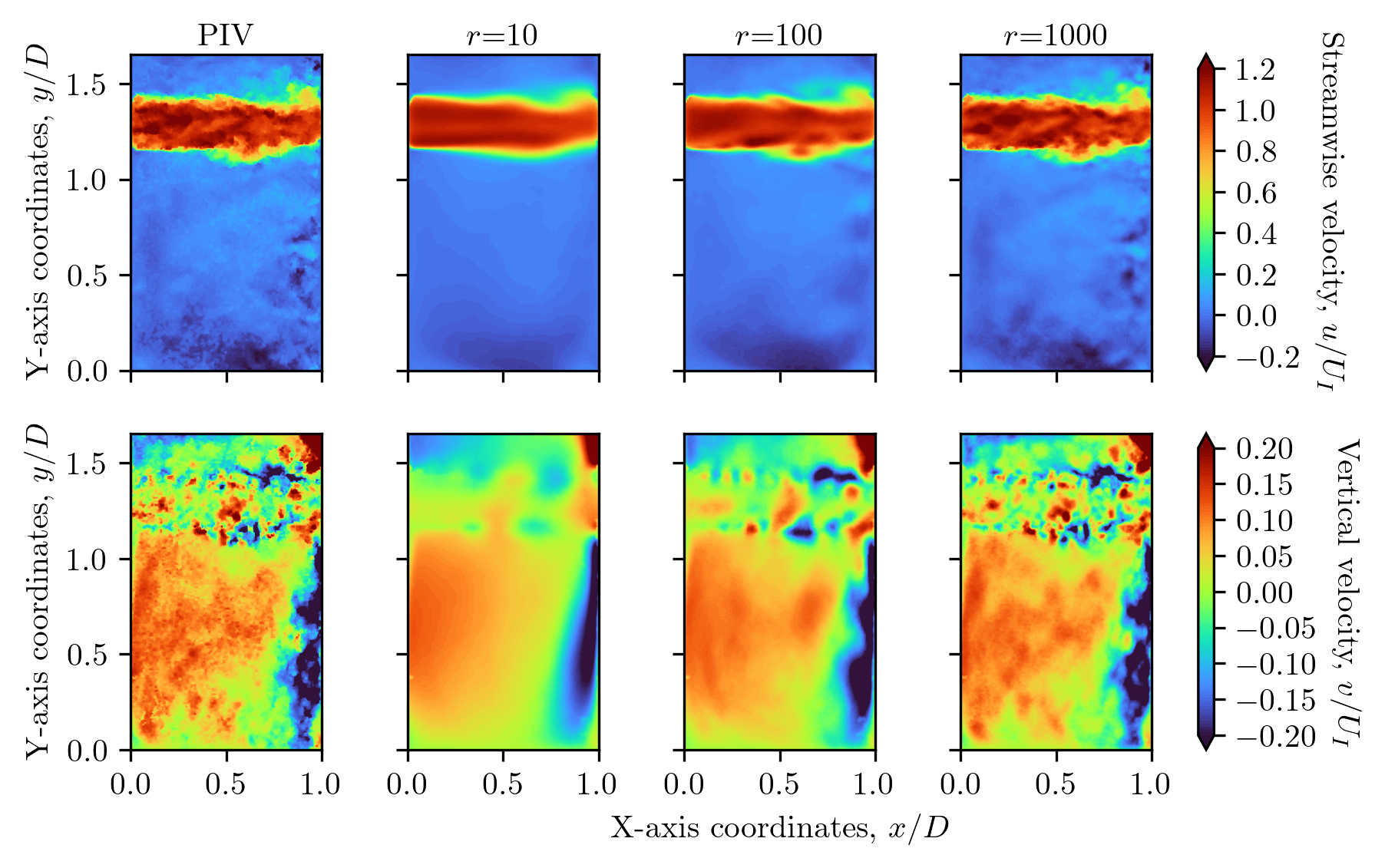} 
    \caption{Reconstructed streamwise velocity and vertical velocity field based on truncated POD modes. $r$ is the rank of the truncated matrix. An animation is shown in Video S3 in online Supplemental Materials}
    \label{fig:POD_ROM}
\end{figure}

Figure \ref{fig:POD_ROM} illustrates instances of reconstructed streamwise velocity and vertical velocity with truncated POD modes of 10, 100, and 1000. This reduced-order dynamics is animated in Video S3. With only 10 POD modes, the ROM already captures key flow features in the HS system such as jet oscillation and vortex street. As the number of modes increases, more detailed flow features are captured. With a higher number of modes, specifically 1000, the ROM yields minimal differences compared with the original data. The reconstruction of the ROM is also computationally efficient, requiring approximately 0.9s on a single CPU processor for the 1000 modes case over 180 s. These results highlight the benefits and real-time computation capabilities of constructing ROM for HS. Once developed, the ROM can be integrated into larger-scale drainage network models as a sub-model for efficient simulation. Unlike simple lumped methods (e.g., CSTR), the ROM preserves a higher-fidelity representation of system dynamics. Moreover, because of the real-time inference capability, ROM can be effectively coupled with control algorithms in the application of adaptive control of treatment systems. However, development based on methods such as POD and SPOD does require data. In this study, direct measurement is available through PIV. In cases where physical modeling is challenging, numerical simulation based on CFD can be a promising solution.

While this study focuses on turbulent flow characteristics in a stormwater treatment system, the PIV technique and modal analysis can also be extended and applied to other water infrastructures, whether for potable water or wastewater treatment. The combination of bench-scale models and PIV measurements facilitates in-depth analysis and visualization of the fundamental physical processes that control treatment system behavior. Further combined with measurement techniques such as planar laser-induced fluorescence (PLIF), the pollutant transport and fate can be directly measured. This not only leads to insights that can guide system design but also generates an extensive database for CFD model development and benchmarking. In the fields of aerospace and mechanical engineering, PIV experiments often serve as a fundamental tool for CFD benchmarking. In contrast, robust application of PIV and open-source measurement databases for water treatment systems in environmental engineering is limited \citep{Catano-Lopera2023, Liu2020a}. This results in many engineering applications where engineers do not have a benchmark database to guide CFD development. There is a significant need and opportunity to extend PIV to other treatment systems and develop and synthesize a robust and open-source measurement database that is widely available to the civil and environmental engineering community.

\section{Conclusion}
This study presents time-resolving, high-resolution particle image velocimetry (PIV) measurements of turbulent flow within a bench-scale hydrodynamic separator (HS), conducted under common surface loading (SLR) conditions. Flow statistics of mean flow and Reynolds stress, are quantified, and their convergence as a function of sampling frequency and duration are examined. Modal analyses using proper orthogonal decomposition (POD) and spectral proper orthogonal decomposition (SPOD) are used to investigate the turbulent flow structures and their spectral characteristics. These analyses elucidated the dominant mechanisms driving flow behavior and offered insights into the dynamics within the system. Additionally, the potential of POD and SPOD for developing reduced-order models (ROM) for water infrastructure is discussed. The results from this study yield the following conclusions:

\begin{itemize}
\item High-resolution PIV is essential for robustly capturing the complex turbulence dynamics in HS. Additionally, PIV measurements can be influenced by the interrogation window size used in the PIV algorithm configuration. While smaller interrogation window sizes (e.g., 16$\times$16) can resolve more detailed flow features, they can also result in reduced correlation and increased measurement uncertainty, as shown in Fig. \ref{fig:PIV_window_size}. Larger window sizes (e.g., 64$\times$64 pixels) similarly increase measurement uncertainty. This uncertainty has a lesser impact on mean flow statistics than on Reynolds stress, as it introduces a degree of artificial filtering. An optimal window size of 32$\times$32 yields the lowest measurement uncertainty. 

\item Sampling frequency and duration play significant roles in obtaining reliable turbulence statistics. Both mean flow statistics and Reynolds stress converge as the sampling frequency and duration increase. To achieve the same level of convergence, Reynolds stress requires a nearly 5x larger sampling size than mean flow statistics, as shown in Fig. \ref{fig:statistics_convergence}. For mean flow statistics, it is more effective to sample over a longer duration with a lower frequency. For higher-order statistics, the advantage of such a sampling strategy is reduced.

\item The analysis of mean flow statistics and Reynolds stresses provides a general characterization of flow patterns within the hydrodynamic separator. The inlet shear layer and impinging jet are significant drivers of turbulence and mixing within the system. In particular, the jet impingement creates high-energy zones along the outlet wall and ultimately contributes to the system bottom, where PM settling occurs, as shown in Fig. \ref{fig:mean_flow}. Modifying these dynamic interactions through strategic separator design has the potential to optimize PM removal efficiencies and minimize the resuspension of settled contaminants, enhancing the overall performance of the HS.

\item The modal analyses of POD and SPOD elucidate the fundamental turbulence structures in the HS. These techniques identified key coherent flow features such as shear layer oscillations and vortex bursting. It is found that 60\% of kinetic energy is captured by the first 126 POD modes and 80\% of kinetic energy is captured by the first 478 POD modes. SPOD, in particular, distinguishes flow features across different frequency spectra, yielding better isolation and characterization of dynamic flow behaviors, as shown in Figs. \ref{fig:SPOD_mode_u} and \ref{fig:SPOD_mode_v}. These modes, once identified, can be effectively used to construct reduced-order models. These reduced-order models can capture the essential dynamics of complex flows in HS using significantly fewer computational resources, making them suitable for real-time control applications and larger-scale drainage network simulations.
\end{itemize}

\section*{Data availability statement}
Some or all data, models, or code generated or used during the study are available in a repository online in accordance with funder data retention policies. The mean flow statistic and Reynolds stress generated from PIV measurements are available at the GitHub repository (\url{https://github.com/Water-Infrastructure-Laboratory/Turbulence-database}).

\section*{Acknowledgment}
This research was supported by the United States Geological Survey (USGS) 104B program through the Tennessee Water Resources Research Center (TNWRRC) at the University of Tennessee, Knoxville, and by the AI Tennessee Initiative at the University of Tennessee. 

\section*{Supplemental materials}
Figs S1-S13 and Video S1-S3 are available online.

\section*{List of symbols}
\begin{tabular}{llll}
$A$ & System bottom area & $\mathbf{A}$ & Coefficient matrix \tabularnewline
$b$ & Snapshots per block in SPOD & $\mathbf{C}$ & Covariance matrix \tabularnewline
$\mathcal{F}$ & Discrete Fourier transform & $D$ & Cylindrical tank diameter \tabularnewline
$D_I$ & Inlet/outlet pipe diameter & $\mathbb{E}$ & Averaging operator \tabularnewline
$\eta$ & Kolmogorov length scale & $Fr_b$ & Bulk Froude number \tabularnewline
$f$ & General frequency & $f_s$ & Sampling frequency \tabularnewline
$g$ & Gravitational acceleration & $\hat{\mathbf{X}}k(f)$ & Fourier coefficients at $f$ for block $k$ \tabularnewline
$N_b$ & Number of blocks in SPOD & $\nu$ & Kinematic viscosity \tabularnewline
$Q$ & Inlet flow rate & $Re_b$ & Bulk Reynolds number \tabularnewline
$Re_I$ & Inlet Reynolds number & $\mathbf{RMD}$ & Relative mean difference \tabularnewline
$\rho_f$ & Fluid density & $S_{u'u'}$ & PSD of streamwise velocity fluctuation \tabularnewline
$S_{v'v'}$ & PSD of vertical velocity fluctuation & $SLR$ & Surface loading rate \tabularnewline
$t_c$ & Convective time unit & $T_d$ & Sampling duration \tabularnewline
$\tau_k$ & Kolmogorov time scale & $\tau_p$ & Particle relaxation time scale \tabularnewline
$U$ & General velocity & $U_I$ & Inlet bulk velocity \tabularnewline
$u$ & X-axis velocity field & $v$ & Y-axis velocity field \tabularnewline
$u'$ & Streamwise velocity fluctuation & $v'$ & Vertical velocity fluctuation \tabularnewline
$\omega_z$ & Z-axis vorticity & $\mathbf{V}(f)$ & Eigenvectors in SPOD at $f$ \tabularnewline
$\mathbf{W}$ & Windowing function matrix & $\mathbf{X}$ & Snapshot data matrix \tabularnewline
$\mathbf{x}_j$ & $j$-th snapshot & $\overline{\mathbf{x}}$ & Time-averaged vector \tabularnewline
$\mathbf{1}_m$ & Column vector of ones & $\boldsymbol{\Lambda}$ & Eigenvalue matrix \tabularnewline
$\boldsymbol{\Lambda}(f)$ & Eigenvalues in SPOD at $f$ & $\boldsymbol{\Phi}$ & Eigenvector matrix \tabularnewline
$\left\langle \cdot \right\rangle$ & Time-averaged quantity & $\left(\cdot\right)^\prime$ & Perturbation from mean \tabularnewline
$\circ$ & Element-wise multiplication & $z$ & Axis perpendicular to paper \tabularnewline
\end{tabular}

\bibliography{filtered_references.bib}
\end{document}